\documentclass[twocolumn,aps,superscriptaddress,longbibliography,amsmath,amssymb,reprint]{revtex4-2} 
\usepackage{bm}
\usepackage{amsmath}
\usepackage{amssymb}
\usepackage{times}
\usepackage[usenames,dvipsnames]{color}
\usepackage[table]{xcolor}
\usepackage{graphicx}
\usepackage{dcolumn}
\usepackage{simplewick}
\usepackage{hyperref}
\usepackage{mathrsfs}
\usepackage{romannum}
\usepackage{dcolumn}
\usepackage[utf8]{inputenc}
\usepackage[T1]{fontenc}
\usepackage{mathptmx}
\usepackage{xr}          
\DeclareUnicodeCharacter{2010}{-}

\hypersetup{
  colorlinks=false,
  colorlinks=true,
  citecolor=blue,
  linkcolor=blue,
  urlcolor=blue}

\begin{document}
\title{Fock-space relativistic coupled-cluster study of systematic shifts and 
       fine-structure constant sensitivity in {\rm Cf$^{17+}$}, {\rm Cm$^{15+}$} 
       and {\rm Bk$^{16+}$} highly charged ions}

\author{Palki Gakkhar}
\affiliation{Department of Physics, Indian Institute of Technology,
             Hauz Khas, New Delhi 110016, India}


\author{D. Angom}
\affiliation{Department of Physics, Manipur University,
             Canchipur 795003, Manipur, India}

\author{B. K. Mani}
\email{bkmani@physics.iitd.ac.in}
\affiliation{Department of Physics, Indian Institute of Technology,
             Hauz Khas, New Delhi 110016, India}

\begin{abstract}

Highly charged ions (HCIs) are among the most promising candidates 
proposed for the development of next-generation optical clocks and 
the precision tests of fundamental physics. 
In this work, we investigate the clock-relevant properties of 
$\rm Cf^{17+}$, $\rm Cm^{15+}$ and $\rm Bk^{16+}$ HCIs using 
Fock-space relativistic coupled-cluster (FSRCC) theory. The corrections 
from Breit interaction, QED effects and perturbative triples are incorporated
to improve the accuracy of results further. 
We compute excitation energies, transition amplitudes, lifetimes, 
hyperfine structure constants, electric quadrupole moments, and 
electric dipole polarizabilities for low-lying states of these ions. 
These properties are further used to characterize important clock 
transitions and estimate key systematic frequency shifts.  
Our calculations reveal that the proposed clock transitions exhibit 
strongly suppressed systematic frequency shifts, highlighting their
suitability for developing accurate optical frequency standards.
We also compute the sensitivity coefficients, $q$,  and enhancement 
factors, $K_{\alpha}$, for the proposed clock transition as a probe to 
variation in the fine-structure constant, $\alpha$. Our calculations predict 
exceptionally large $K_{\alpha}$ of $-55.3$ and $57.5$ for 
$5f_{5/2} \rightarrow 6p_{1/2}$ transition in $\rm Cf^{17+}$ and 
$6p_{1/2} \rightarrow 5f_{5/2}$ transition in $\rm Bk^{16+}$, respectively. 
These giant $K_{\alpha}$s place the proposed transitions among the most 
sensitive optical probes for the laboratory searches for $\alpha$ 
variation. Further, a comparative analysis of proposed transitions 
in $\rm Cm^{15+}$ and $\rm Bk^{16+}$ shows that, $5f_{5/2} \rightarrow 5f_{7/2}$ 
transition possess smaller systematic shifts and is therefore 
attractive for precision frequency metrology, whereas $6p_{1/2} \rightarrow 5f_{5/2}$ 
transition exhibit significantly enhanced sensitivity to $\alpha$ variation.
\end{abstract}

\maketitle

{\em Introduction--}The extraordinary accuracy and stability of optical atomic 
clocks have enabled applications ranging from precision frequency metrology and relativistic 
geodesy \cite{mehlstaubler_18} to the test of general relativity \cite{kolkowitz-16, 
bondarescu-12} and the searches for physics beyond the Standard Model \cite{dzuba-18, berengut-18}. 
Among the key applications, 
the search for possible temporal variation of fine-structure constant, 
$\alpha$, has attracted a considerable interest \cite{safronova-19, prestage-95, karshenboim-10}. 
Any detectable variation in $\alpha$ would provide evidence for physics 
beyond the Standard Model and can offer important insights into cosmology and the nature 
of dark matter. The sensitivity of a clock-comparison experiment is determined
by the difference in the enhancement factors ($K_{\alpha}$) of the two clock 
transitions. For example, comparisons between $^{199}$Hg$^{+}$ and $^{27}$Al$^{+}$ 
optical clocks, for which $\Delta K_{\alpha}\approx 2.95$, constrained the 
temporal variation of $\alpha$ at the level of $10^{-17}$ yr$^{-1}$\cite{rosenband-08}. 
Further, recent measurements of the frequency ratio between E2 and E3 clock 
transitions in $^{171}$Yb$^{+}$, characterized by $\Delta K_{\alpha}\approx -6.95$,
improved this limit to $10^{-18}$ yr$^{-1}$\cite{lange-21}. These remarkable 
achievements have motivated the search for atomic systems possessing 
substantially larger $K_{\alpha}$s.

Highly charged ions (HCIs) have emerged as promising candidates for 
probing $\alpha$ \cite{berengut-12,yu-23,safronova-14}. Their compact electronic 
structure leads to reduced sensitivities to external perturbations. In addition, 
the large nuclear charge, strong relativistic effects, and high ionization 
potentials in HCIs enhance their sensitivity to $\alpha$ variation. 
Among the proposed HCI based clocks, ions near the $5f$-$6p$ level 
crossing are of particular interest \cite{berengut-12a}. 
One of the most prominent examples is $\rm Cf^{17+}$, which was identified 
by Berengut \textit{et al.} \cite{berengut-12} as a highly promising 
candidate for probing variation in $\alpha$. The $5f_{5/2} \rightarrow 6p_{1/2}$ 
optical transition in this ion combines a narrow linewidth with one of the 
largest enhancement factors predicted for any optical-clock transition. 
Moreover recently, Porsev \textit{et al.} \cite{porsev-20,porsev-26} carried out 
high-precision coupled-cluster (CC) calculations and demonstrated that 
$\rm Cf^{17+}$ possesses favorable clock characteristics together with 
extraordinary sensitivity to $\alpha$ variation.
The neighboring ions $\rm Cm^{15+}$ and $\rm Bk^{16+}$ have also been proposed as 
attractive clock candidates by Dzuba \textit{et al.} \cite{dzuba-24}. 
The states $6p_{1/2}$, $5f_{5/2}$, and $5f_{7/2}$ in these HCIs lie within 
the optical domain and give rise to two potential clock transitions,
$6p_{1/2} \rightarrow 5f_{5/2}$ and $5f_{5/2}\rightarrow5f_{7/2}$. 
The former one involves orbitals of markedly different relativistic character, 
whereas the latter corresponds to a fine-structure transition within the $5f$ 
manifold. These contrasting electronic configurations in $\rm Cm^{15+}$ and
$\rm Bk^{16+}$ make them interesting systems for investigating the 
interplay between clock accuracy and sensitivity to the variation 
in fundamental constants.

Despite their considerable promise, experimental information on the low-lying 
spectra and clock-relevant properties of these HCIs remains extremely limited. 
Consequently, the characterization of these systems presently relies almost 
entirely on theoretical predictions. However, the available theoretical 
studies are also limited. For $\rm Cf^{17+}$, only the pioneering 
work of Berengut \textit{et al.} \cite{berengut-12} and the subsequent CC 
studies of Porsev \textit{et al.} \cite{porsev-20,porsev-26} are available, 
whereas for $\rm Cm^{15+}$ and $\rm Bk^{16+}$ the available data are largely 
restricted to the recent linearized CC calculations of Dzuba \textit{et al.} \cite{dzuba-24}.
Although these works established the exceptional potential of these ions, a 
comprehensive and uniform investigation of their clock-relevant properties is still 
lacking. In particular, while both $6p_{1/2} \rightarrow 5f_{5/2}$ and 
$5f_{5/2} \rightarrow 5f_{7/2}$ transitions were proposed as promising 
transitions in $\rm Cm^{15+}$ and $\rm Bk^{16+}$, a quantitative comparison 
of their relative merits has not yet been carried out. Importantly, key 
properties such as transition amplitudes, lifetimes, hyperfine structure 
constants, electric quadrupole moments, dipole polarizabilities, and 
systematic frequency shifts have not been investigated consistently within 
a common high-precision many-body framework which systematically 
incorporates correlation and relativistic effects to the same level 
of accuracy. Moreover, the calculations reported for $\rm Cm^{15+}$ 
and $\rm Bk^{16+}$ employ a linearized CC treatment, missing the 
important correlation effects from nonlinear terms.

In this work, we perform a systematic Fock-space relativistic coupled-
cluster (FSRCC) \cite{mani-09, mani-10, mani-17} investigation of the 
spectroscopic properties of 
low-lying states of $\rm Cf^{17+}$, $\rm Cm^{15+}$, and $\rm Bk^{16+}$ HCIs. 
Electron-correlation effects are treated to all orders within the CC 
framework, while the corrections from the Breit interaction, QED effects 
and perturbative triples are incorporated to improve the accuracy of
calculations further. We calculate the aforementioned clock's relevant 
spectroscopic properties. Further, we estimate the dominant systematic 
shifts in clock transition frequencies, including blackbody-radiation (BBR), 
electric-quadrupole (EQ), quadratic Zeeman, and micromotion-induced (MI). 
In addition, relativistic sensitivity coefficients and enhancement factors 
are determined as a probe to $\alpha$ variation. 
Our work also enables a direct comparison of the competing clock transitions 
in $\rm Cm^{15+}$ and $\rm Bk^{16+}$, providing a quantitative assessment of 
their suitability for high-accuracy clocks and as a probe 
for $\alpha$ variation. 

\begin{figure*}
\includegraphics[scale = 0.28, angle=0]{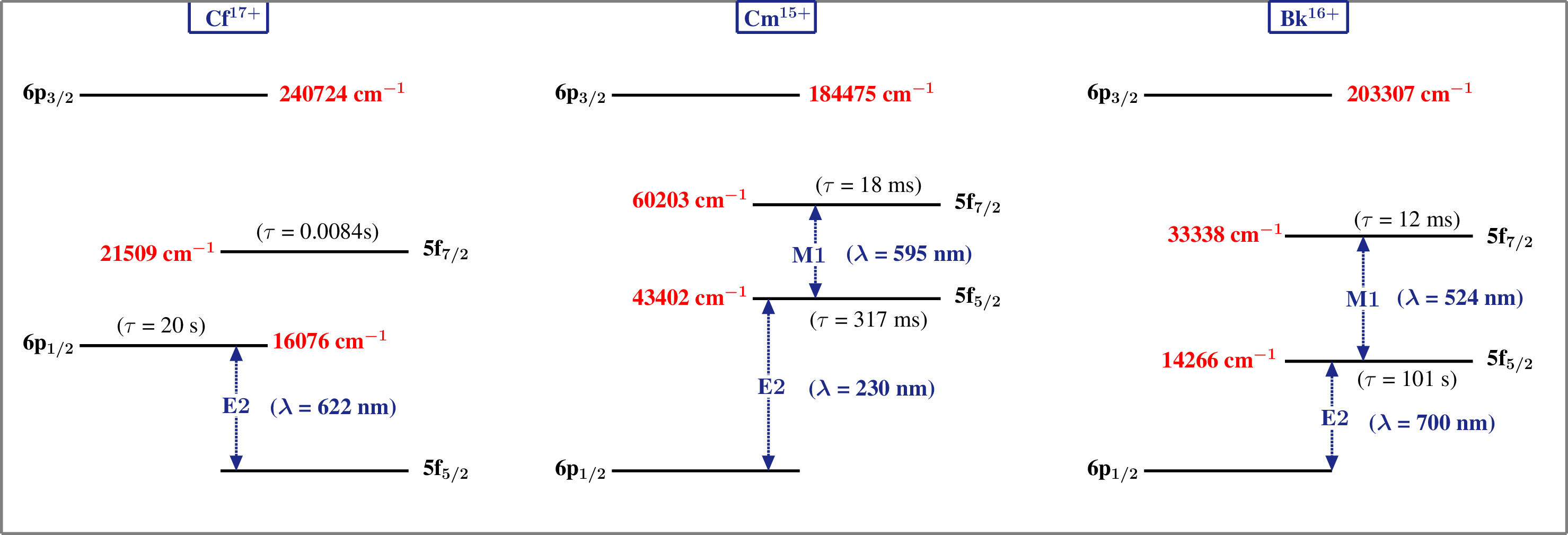}
\caption{Energy level diagrams of low-lying states of $\rm Cf^{17+}$, $\rm 
Cm^{15+}$ and $\rm Bk^{16+}$ HCIs. Energies, lifetimes and wavelengths shown in the 
figures are obtained from present calculations. The proposed clock transitions 
are indicated by double-headed arrows.}
\label{energy_levels}
\end{figure*}


{\em Basis Set, Excitation Energy, Lifetime and Hyperfine Constants--}We have 
used Gaussian type orbitals (GTOs) as basis function for FSRCC calculations \cite{mohanty-91}.  
The self-energy (SE) correction to orbitals' energies is incorporated through 
the model Lamb-shift operator \cite{shabaev-15}, whereas the correction from 
the vacuum polarization (VP) is accounted using Uehling potential \cite{uehling-35}.
The GTO parameters are optimized to reproduce the self-consistent field (SCF)
and single-particle energies obtained from B-spline \cite{oleg-16}.
Tables \ref{basis} and \ref{conv-basis} in Supplemental Material (SM) lists 
the optimized GTO parameters and comparison of GTO energies with B-spline data.
We observe an excellent agreement for both single-particle as well as SCF 
energies with B-spline results.
Further, Fig. \ref{conv-diag} in SM demonstrates the convergence of computed 
properties with basis size.

Table \ref{energy} lists the excitation energies (EEs) for low-lying excited states
of considered HCIs. As to be expected, Dirac-Fock (DF) constitutes the dominant
part of EE in all three ions.  For example, it accounts for approximately 53\%,
118\%, and 155\% of total EE for first excited state in all three HCIs.
The second largest contribution comes from the electron correlation due 
to residual Coulomb interaction. This exhibits a pronounced state dependence 
both in magnitude as well as sign due to different strengths 
of {\em core-valence} correlation effects. The largest cumulative 
Breit+QED contribution is observed to be $\approx$ 1.3\% in $6p_{1/2}$ 
state of $\rm Cf^{17+}$ (Fig. \ref{energy_bqed} in SM). This is a significant contribution 
and essential for high accuracy clocks.
To the best of our knowledge, there are no reported values of experimental EEs for HCIs considered.
In the case of $\rm Cf^{17+}$, for $6p_{1/2}$, there is a noticeable 
difference among previous results. The lowest and highest reported values
differ by approximately 13\%. Though Refs. \cite{porsev-20, porsev-26}
employ the same CCSDT method, their results differ by about 4\%. 
Our EE for $6p_{1/2}$ is lower than all previous values. 
The lower value, compared to Ref. \cite{berengut-12}, could be attributed to 
the differences in the computational methodologies employed in the two works, 
whereas the difference with Ref. \cite{porsev-26} may partly be attributed 
to different implementations of Breit interaction.
For $6p_{3/2}$ and $5f_{7/2}$ states, our results are consistent
with previous results and, as to be expected, more closer to the CCSDT
results from Refs. \cite{porsev-20, porsev-26}.
The observed differences among previous results and ours can be
attributed to the varying treatment of electron-correlation and
relativistic effects.
For $\rm Cm^{15+}$ and $\rm Bk^{16+}$, we observed large differences in the
EEs for first two excited states compared to linearized coupled-cluster theory
with single and double approximation (LCCSD) results of 
Ref. \cite{dzuba-24}. For example, for $\rm Cm^{15+}$, our results
are larger by $\approx$ 13\% and 10\%, respectively. For $\rm Bk^{16+}$, the
difference is even more.
The reason could be attributed to the different treatment of electron 
correlations in the two works. In Ref. \cite{dzuba-24}, {\em core-core} 
and {\em core-valence} correlations are treated using a linearized 
CC theory, whereas the present work uses a non-linear CC theory. 
Additional differences could also arise because of different 
implementation of Breit and QED effects.

\begin{table}
    \caption{Excitation energies (cm$^{-1}$) of low-lying states of Cf$^{17+}$,
	Cm$^{15+}$ and Bk$^{16+}$ HCIs using FSRCC theory. For a quantitative
	assessment of electron correlation, separate contributions from
	different correlation terms are also provided. Our final values
	are compared with other available theoretical results.}
  \label{energy}
  \begin{ruledtabular}
  \begin{tabular}{lllllc}
    States &  {\textrm{DF}}  & {\textrm{FSRCC}} & Breit+ & {\textrm{Total}} &  Others  \\
           &                 &                  & QED &  &  \\
      \hline
      \multicolumn{6}{c}{$\rm Cf^{17+}$} \\
      $6p_{1/2}$ & 8450       & 16288   & $-212$ & 16076   & $18686^{\rm a}$, $20611^{\rm b}$,\\
       &  &  & &   &  $21456^{\rm c}$  \\
      $6p_{3/2}$ &  233476    & 240772  & $-48$  & 240724  & $242811^{\rm a}$, $243081^{\rm b}$, \\
       &  &  & &   & $244125^{\rm c}$ \\
      $5f_{7/2}$ &  20448     &  21516  & $-7$   & 21509   & $21848^{\rm a}$, $20895^{\rm b}$,\\
       &  &  & &   & $20870^{\rm c}$ \\
      \hline
      \multicolumn{6}{c}{$\rm Cm^{15+}$} \\
      $5f_{5/2}$ &  51258   & 43302    & 100  & 43402  &  $38375^{\rm d}$   \\
      $5f_{7/2}$ &  67020   & 60101    & 102  & 60203  &  $54730^{\rm d}$   \\
      $6p_{3/2}$ &  183280  & 184398   &  77  & 184475 & $182359^{\rm d}$    \\
      \hline
      \multicolumn{6}{c}{$\rm Bk^{16+}$} \\
      $5f_{5/2}$ &  22176  &  14139   &  127 & 14266  &  $8880^{\rm d}$   \\
      $5f_{7/2}$ &  40208  &  33221   &  117 & 33338  &  $27520^{\rm d}$   \\
      $6p_{3/2}$ &  203445 & 203201   &  106 & 203307 & $202403^{\rm d}$    \\
        \end{tabular}
        \end{ruledtabular}
        \begin{tabbing}
$^{\rm a}$Ref.\cite{berengut-12}[CI + MBPT],
$^{\rm b}$Ref.\cite{porsev-20}[CCSDT],
$^{\rm c}$Ref.\cite{porsev-26}[CCSDT], \\
$^{\rm d}$Ref.\cite{dzuba-24}[LCCSD].
\end{tabbing}
\end{table}


Table \ref{lifetime} presents our results on
transition rates ($A$) and lifetimes ($\tau$) for some low-lying excited
states of HCIs considered.
For a quantitative assessment, we have explicitly given the relevant decay
channels and corresponding contributions. To the best
of our knowledge, there are no experimental results on $\tau$ for 
comparison for any of the considered HCIs. Our calculated $\tau$ for $6p_{1/2}$ 
state of $\rm Cf^{17+}$ is observed to be approximately three times
larger than that of only available result \cite{porsev-20}.
The reason for this could be attributed to the difference in EEs 
in the two works.  
In contrast, the EEs of $6p_{3/2}$ and $5f_{7/2}$ states are in closer 
agreement with Ref. \cite{porsev-20}, resulting in good agreement 
between the corresponding $\tau$s.
A similar trend is also observed for $\rm Cm^{15+}$ and $\rm Bk^{16+}$ ions,
where we observe a noticeable difference for first excited state compared to
LCCSD results of Ref. \cite{dzuba-24}. This difference could be attributed
to the more accurate treatment of {\em core-core} and {\em core-valence}
correlations in our calculations.
The contributions from Breit and QED corrections are observed to be not very 
significant in all three ions (Fig \ref{lifetime_bqed} in SM). The largest Breit+QED contribution 
is observed to be $-0.02$\% in the case of $6p_{1/2}$ state of $\rm Cf^{17+}$.
Whereas, perturbative triples are observed to have non negligible contributions, 
with a largest contribution of 0.3\% in the case of $6p_{1/2}$ state 
of $\rm Cf^{17+}$.

\begin{table}
\begin{center}
\caption{FSRCC results for transition rates ($A$) and lifetimes ($\tau$) for
	low lying states of {\rm Cf$^{17+}$}, {\rm Cm$^{15+}$}
	and {\rm Bk$^{16+}$} HCIs. The values given for $\tau$ from this work 
	also include the contribution from Breit, QED and perturbative 
	triples corrections.}
\begin{ruledtabular}
\begin{tabular}{lllrrrrrr}
  Upper & Lower & Decay & A (s$^{-1}$) & \multicolumn{2}{c}{\rm $\tau$ (s)} \\
  \cline{5-6}
    state & state & channel &  & This work & Other cal.  \\
  \hline
  \multicolumn{6}{c}{\rm Cf$^{17+}$} \\
$6p_{1/2}$ & $5f_{5/2}$ & E2 &  5.115[$-2$] & 19.607 & $6.0^{\rm a}$\\
$6p_{3/2}$ & $6p_{1/2}$ & E2 &  2.825[4]  &  6.7[$-6$] & $7.0[-6]^{\rm a}$\\
           &            & M1 &  9.746[4]  &  & \\
           & $5f_{5/2}$ & E2 &  4.715[3]  &  &\\
           &            & M1 &  4.396[$-1$] &  &\\
           & $5f_{7/2}$ & E2 &  1.894[4]  &   &\\
$5f_{7/2}$ & $5f_{5/2}$ & E2 &  7.427[$-3$] & 0.0084 & $0.0095^{\rm a}$\\
           &            & M1 &  1.188[2]  &  &\\
\hline
  \multicolumn{6}{c}{\rm Cm$^{15+}$} \\
$5f_{5/2}$ & $6p_{1/2}$ & E2 & 3.158     &  0.318 & $0.576^{\rm b}$\\
$5f_{7/2}$ & $5f_{5/2}$ & E2 & 2.936[$-3$] & 0.018 & $0.02^{\rm b}$\\
           &            & M1 & 5.653[1]  &   &\\
$6p_{3/2}$ & $6p_{1/2}$ & E2 & 1.320[4]  & 1.5[$-5$]&\\
           &            & M1 & 5.352[4]  &   &\\
           & $5f_{5/2}$ & E2 & 4.476[2]  &   &\\
           &            & M1 & 9.561[$-2$] &  &\\
           & $5f_{7/2}$ & E2 & 1.506[3]  &   &\\
\hline
  \multicolumn{6}{c}{\rm Bk$^{16+}$} \\
$5f_{5/2}$ & $6p_{1/2}$ & E2 & 9.874[$-3$] & 101.388 & $787^{\rm b}$\\
$5f_{7/2}$ & $5f_{5/2}$ & E2 & 4.731[$-3$] &  0.012 & $0.013^{\rm b}$\\
           &            & M1 & 8.284[1]  &    &\\
$6p_{3/2}$ & $6p_{1/2}$ & E2 & 1.948[4]  & 9.9[$-6$] &\\
           &            & M1 & 7.271[4]  &  &\\
           & $5f_{5/2}$ & E2 & 1.679[3]  &  &\\
           &            & M1 & 2.237[$-1$] &  &\\
           & $5f_{7/2}$ & E2 & 6.310[3]  &  &\\
 \end{tabular}
\end{ruledtabular}
\begin{tabbing}
$^{\rm a}$Ref.\cite{porsev-20}- CCSDT,
$^{\rm b}$Ref.\cite{dzuba-24}- LCCSD
\end{tabbing}
\label{lifetime}
\end{center}
\end{table}

Table \ref{mhfs_table} presents our results on magnetic-dipole ($A/g_I$)
and electric-quadrupole ($B/Q$) hyperfine structure (HFS) constants for
low-lying states of HCIs.
For $A/g_I$, our calculations show that the contributions from Breit,
SE, and perturbative triples corrections are not very significant 
in all three ions (Fig. \ref{mhfs_bqed} in SM). Whereas, VP has a noticeable 
contribution of approximately 0.4\% to $6p_{1/2}$ state in all three ions. 
For $B/Q$, we get negligible contributions from all higher-order relativistic 
and correlation corrections.
To the best of our knowledge, no experimental measurements of $A$ or $B$
are available for any of the states considered. Theoretical calculations 
are too very few.
Our calculated $A/g_{I}$ for $5f_{5/2}$ state of Cf$^{17+}$ is approximately 7.4\%
smaller than Ref. \cite{porsev-20}, whereas for $6p_{1/2}$ state, the two
calculations are closer. Since Ref. \cite{porsev-20} does not provide
separate contributions from higher-order corrections, a detailed term-by-term
comparison is not feasible. It should, however, be noted that the uncertainty
related to HFS constants in Ref. \cite{porsev-20} is estimated to be 20-30\%.
Comparing our results for $\rm Cm^{15+}$ and $\rm Bk^{16+}$, there are
differences with only available calculation \cite{dzuba-24} using LCCSD
theory. And, as mentioned earlier, the reason for this could be the missing
non-linear correlation effects in Ref. \cite{dzuba-24}, along with small 
differences Briet and QED corrections.
For $B/Q$, our calculations provide the first theoretical data for the
HCIs considered. Since both $A$ and $B$s serve as a probe for electronic
wavefunction in nucleus region, they constitute a stringent test for 
the accuracy of many-body wavefunctions in HCIs.

\begin{table}[h]
	\caption{$A/g_{I}$($B/Q$) in the units of MHz/$ \mu_{N}$(MHz/eb) for
	low lying states of considered HCIs. The cumulative Breit, QED and perturbative 
	triples corrections are listed separately. Values from previous works
	correspond to $A/g_I$.}
  \label{mhfs_table}
  \begin{ruledtabular}
  \begin{tabular}{rrrrr}
      States & {\textrm{FSRCC}} & {\textrm{Breit+QED}} & {\textrm{Total}}&Others \\
             &      & {\textrm{+Triples}} & &  \\
      \hline
      \multicolumn{5}{c}{$\rm Cf^{17+}$} \\
	  $5f_{5/2}$ &   1764(3295)   &$-4$(9)     &  1760(3304)  & $1900^{\rm a}$    \\
	  $6p_{1/2}$ &  198766(0)  & $791$(0)            &199557(0)  & $195000^{\rm a}$   \\
	  $6p_{3/2}$ &  10450(26362)  & $-8$(17)   & 19442(26379)  &          \\
	  $5f_{7/2}$ &  1231(3857)    &   3(11)    &  1234(3868) &           \\
      \hline
      \multicolumn{5}{c}{$\rm Cm^{15+}$} \\
	  $6p_{1/2}$ &157685(0)       & $633$(0)    &  158314(0)  & $154000^{\rm b}$  \\
	  $5f_{5/2}$ &    1392(2381)  &   9(6)      &    1401(2387)  & $1550^{\rm b}$   \\
	  $5f_{7/2}$ &    1011(3324)  & $2$(12)     &    1013(3336)  & $689^{\rm b}$   \\
	  $6p_{3/2}$ &    8821(24296) & $-7$(17)    &    8814(22120)  &          \\
      \hline
      \multicolumn{5}{c}{$\rm Bk^{16+}$} \\
	  $6p_{1/2}$ & 178846(0)  & $712$(0)   &  179558(0)  & $174000^{\rm b}$   \\
	  $5f_{5/2}$ &   1568(3138)  &   8(6)     &    1576(3144)  &   $1740^{\rm b}$   \\
	  $5f_{7/2}$ &   1120(3673)  & $0$(10)     &    1120(3683)  &    $772^{\rm b}$   \\
	  $6p_{3/2}$ &   9644(24296)  & $-6$(18)    &   9638(24314)  &           \\
        \end{tabular}
        \end{ruledtabular}
        \begin{tabbing}
$^{\rm a}$Ref.\cite{porsev-20}[CCSDT],
$^{\rm b}$Ref.\cite{dzuba-24}[LCCSD]. \\
\end{tabbing}
\end{table}


{\em Systematic shifts--} Now we present and discuss our results on leading
frequency shifts for proposed clock transitions and assess their suitability
for high-precision optical clock applications.
Within the static approximation, the BBR-induced frequency shift for 
a clock transition between $|g\rangle$ and $|e\rangle$ states at 
temperature $T$ is given by
\begin{equation}
\Delta \nu_{BBR}(T) = -\frac{1}{2h}\langle E^2(T)\rangle \Delta\alpha_s,
\label{BBR}
\end{equation}
where $\Delta\alpha_s$ is the differential static scalar electric
dipole polarizability between $|g\rangle$ and $|e\rangle$ states.
$\langle E^{2} (T)\rangle = (831.9 V/m)^{2} \left(\frac{T}{300 K}\right)^{4}$ is the
mean-squared electric field associated with blackbody radiation and $h$ is
the Planck's constant. As it is evident from Eq. (\ref{BBR}), the accuracy
of $\Delta \nu_{BBR}(T)$ is governed by the accuracy of $\Delta\alpha_s$.

Table \ref{static-pol} in SM lists $\alpha_s$ for relevant states of considered
HCIs. The dominant contribution is observed to originate from the closed-shell 
core. For example, in the case of $\rm Cf^{17+}$, the core contributes 
approximately 82\%, 94\%, and 94\% of total $\alpha_s$ for $6p_{1/2}$, 
$5f_{5/2}$, and $5f_{7/2}$ states, respectively.
Similar trends are also observed for $\rm Cm^{15+}$ and $\rm Bk^{16+}$ ions.
Among the valence-electrons, $6d_{3/2}$ state provides the largest contribution 
of approximately 17-18\% to $6p_{1/2}$.
For $5f_{5/2}$ and $5f_{7/2}$, the dominant valence contributions arise 
mainly from $6d_{3/2}$ and $5g$ states and contribute only a few percent 
to total $\alpha_s$. The Briet and QED corrections are found to have 
a negligible influence on $\alpha_s$. 
In contrast, perturbative triples are observed to have more
significant contributions, with the largest one of 0.3\%.

To the best of our knowledge, no experimental measurements of $\alpha_s$
are available for any of the states considered. Theoretical data are also too 
scarce. Our calculated $\alpha_s$ for $6p_{1/2}$ state of $\rm Cf^{17+}$
is approximately 18\% larger than Ref. \cite{porsev-20}. The $\alpha_s$ for
$5f_{5/2}$ state from two works are closer with a small difference of about 2\%.
For $\rm Cm^{15+}$ and $\rm Bk^{16+}$, the only available theoretical results 
are those in Ref. \cite{dzuba-24}. Relative to this, the present calculations
differ by approximately 13\%, 4\%, and 4\% for $6p_{1/2}$, $5f_{5/2}$, and
$5f_{7/2}$ states, respectively. These differences could be attributed
to more complete treatment of electron correlation in our work through
nonlinear contributions in CC.

Using our calculated $\Delta\alpha_s$, we estimate the BBR shifts associated
with proposed clock transitions in considered HCIs (Table \ref{bbr_shift}).
For $5f_{5/2} \rightarrow 6p_{1/2}$ clock transition in $\rm Cf^{17+}$, 
our estimated $\delta \nu_{BBR}/\nu$, is approximately an order of magnitude 
larger than the value in Ref. \cite{porsev-20}.
This difference arises primarily due to the differences in $\Delta\alpha_s$ 
used in the two calculations.
The same trend is observed for $6p_{1/2} \rightarrow 5f_{5/2}$ clock 
transition in $\rm Cm^{15+}$ with respect to Ref. \cite{dzuba-24}.
In contrast, $\delta \nu_{BBR}/\nu$ for $5f_{5/2} \rightarrow 5f_{7/2}$ 
transition in $\rm Cm^{15+}$ is of the same order as that in Ref. \cite{dzuba-24}.
For $\rm Bk^{16+}$, our estimated $\delta \nu_{BBR}/\nu$ for 
$5f_{5/2} \rightarrow 5f_{7/2}$ transition is an order smaller 
than in Ref. \cite{dzuba-24}.
Overall, our analysis indicates that $5f_{5/2} \rightarrow 5f_{7/2}$ transition
exhibit smaller $\delta \nu_{BBR}/\nu$ than $6p_{1/2} \rightarrow 5f_{5/2}$ 
transition in $\rm Cm^{15+}$ and $\rm Bk^{16+}$. This is due to extremely small $\Delta\alpha_s$ 
associated with fine-structure $5f_{5/2} \rightarrow 5f_{7/2}$ transitions.


\begin{table}
    \caption{Relative frequency shift due to BBR in $\rm Cf^{17+}$,
    $\rm Cm^{15+}$ and $\rm Bk^{16+}$ HCIs.}
  \label{bbr_shift}
  \begin{ruledtabular}
  \begin{tabular}{ccccc}
      Ion & {\textrm{Transition}} & {\textrm{$\Delta \alpha_s$ (a.u.)}}
	  & {\textrm{$\delta \nu_{BBR}/\nu$}} & Others  \\
      \hline
      $\rm Cf^{17+}$ &  $5f_{5/2} - 6p_{1/2}$ &  0.1418     &  $-2.5 \times 10^{-18}$
                     &  $5.9 \times 10^{-19}$ [a]   \\
      $\rm Cm^{15+}$ &  $6p_{1/2} - 5f_{5/2}$ &  $-0.1595$  &  $1.1 \times 10^{-18}$
                     &  $-3.4 \times 10^{-19}$ [b]   \\
                     &  $5f_{5/2} - 5f_{7/2}$ &  $-0.0045$  &  $7.8 \times 10^{-20}$
                     &  $2.2 \times 10^{-20}$ [b]      \\
      $\rm Bk^{16+}$ &  $6p_{1/2} - 5f_{5/2}$ &  $-0.1545$  &  $3.2 \times 10^{-18}$
                     &  $-1.5 \times 10^{-18}$ [b]      \\
                     &  $5f_{5/2} - 5f_{7/2}$ &  $-0.0005$  &  $7.6 \times 10^{-21}$
                     &  $3.4 \times 10^{-20}$ [b]         \\
        \end{tabular}
        \end{ruledtabular}
        \begin{tabbing}
${\rm [a]}$Ref.\cite{porsev-20}[CCSDT],
${\rm [b]}$Ref.\cite{dzuba-24}[LCCSD]. \\
\end{tabbing}
\end{table}


Next, we estimate the EQ shift which originates
due to an interaction between atomic quadrupole moment and an external 
electric-field gradient. In trapped-ion optical clocks, electric-field 
gradients generated by the trapping potential can perturb states 
with non-vanishing quadrupole moments, leading to shifts in 
the clock frequency.
The interaction of electric quadrupole moment ($\Theta$) with
an external electric-field gradient gives rise to a frequency shift
\begin{equation}
  \Delta \nu_Q = -\frac{1}{2h}\Delta\langle Q_0\rangle
  \frac{\partial \mathscr{E}_z}{\partial z},
\end{equation}
where $Q_0 = \sum_i\left(\frac{3}{2}z_i^2-\frac{1}{2}r_i^2\right)$ is
the zero component of the electric quadrupole operator,
$\Delta\langle Q_0\rangle$ denotes the difference in the
expectation values of $Q_0$ between $|g\rangle$ and $|e\rangle$ states, 
and $\Theta = 2\langle \gamma J,M_J=J|Q_0|\gamma J,M_J=J\rangle$.

\begin{table}
    \caption{The electric quadrupole moment (|e| a$_{0}^{2}$) for low lying
	states of $\rm Cf^{17+}$, $\rm Cm^{15+}$ and $\rm Bk^{16+}$ HCIs.}
  \label{quadrupole_moment}
  \begin{ruledtabular}
  \begin{tabular}{cccccccc}
      States & \multicolumn{2}{c}{$\rm Cf^{17+}$} & \multicolumn{2}{c}{$\rm Cm^{15+}$} &
      \multicolumn{2}{c}{$\rm Bk^{16+}$} \\
      \cline{2-3}   \cline{4-5}  \cline{6-7}
      & Present & Others & Present & Others & Present & Others \\
      \hline

      $5f_{5/2}$ & $-0.398$  &  $0.39^{\rm a}$  &  $-0.464$ & $-0.459^{\rm b}$ &
                   $-0.428$  &  $-0.507^{\rm b}$   \\
      $5f_{7/2}$ & $-0.476$  &                  &  $-0.556$ & $-0.550^{\rm b}$ &
                   $-0.512$  &  $-0.604^{\rm b}$    \\
      $6p_{3/2}$ & $-0.666$  &                  &  $-0.754$ &          &
                   $-0.708$  &            \\

        \end{tabular}
        \end{ruledtabular}
        \begin{tabbing}

$^{\rm a}$Ref.\cite{porsev-20}-CCSDT, 
$^{\rm b}$Ref.\cite{dzuba-24}-LCCSD, \\

\end{tabbing}
\end{table}

The $\Theta$ from our calculations for low-lying states are listed
in Table \ref{quadrupole_moment}.
For Cf$^{17+}$, our calculated $\Theta$ for $5f_{5/2}$ state is in
excellent agreement with the value reported in Ref. \cite{porsev-20},
except the sign difference which could arise due to different conventions 
used in the definition of single-electron quadrupole operator.
For Cm$^{15+}$ and Bk$^{16+}$, our results slightly differ from those reported 
in Ref. \cite{dzuba-24} due to the reasons discussed earlier.
Since EQ shift is directly proportional to the differential $\Theta$, 
$\Theta$ provides a measure of the intrinsic sensitivity of a clock 
transition to the electric-field gradients.
The differential $\Theta$ are calculated to be 0.398 |e|a$_{0}^{2}$ for 
the $5f_{5/2} \rightarrow 6p_{1/2}$ transition in $\rm Cf^{17+}$ and 
0.464 |e|a$_{0}^{2}$ and 0.428 |e|a$_{0}^{2}$ for $6p_{1/2} \rightarrow 5f_{5/2}$ 
transitions in Cm$^{15+}$ and Bk$^{16+}$, respectively.
In contrast, for $5f_{5/2} \rightarrow 5f_{7/2}$ transition, since the 
two $5f$ states have very close $\Theta$, differential $\Theta$ is very small; 
0.092 |e|a$_{0}^{2}$, 0.084 |e|a$_{0}^{2}$ for Cm$^{15+}$ and Bk$^{16+}$, 
respectively. These values are approximately five times smaller
than those associated with $6p_{1/2} \rightarrow 5f_{5/2}$ transitions.
Therefore, from the perspective of EQ systematic effects, the
$5f_{5/2} \rightarrow 5f_{7/2}$ transition is expected to provide a more robust
clock reference than $6p_{1/2} \rightarrow 5f_{5/2}$ transition in
both Cm$^{15+}$ and Bk$^{16+}$.
However, the EQ shift can be eliminated for clock transition in all
three systems by choosing the $F=3$, $m_F= \pm 2$ hyperfine sublevels, for 
which the angular factor vanishes, irrespective of the magnitude of the 
electronic quadrupole moments.


We now assess the micromotion-induced frequency shifts.
In radio-frequency (RF) Paul trap, the oscillating trapping field induces a driven
motion of confined ion, commonly referred to as the micromotion. This gives rise
to two competing contributions to the clock frequency: the second-order
Doppler shift associated with the ion velocity and, an AC Stark shift caused
by oscillating electric field. These two effects can be made to
cancel each other at a particular RF drive frequency, known as the magic trap
drive frequency. Operation of trap at this frequency suppresses the net
micromotion-induced  frequency shift and hence improves the attainable
clock accuracy. The magic trap drive frequency can be expressed as
\begin{equation}
    \Omega = \frac{|e|}{M_{i}c}\sqrt{-\frac{h \nu_{\rm clock}}{\Delta \alpha_s}},
    \label{magic trap}
\end{equation}
where M$_{i}$ $\approx A m_{p}$ is the ion mass ($m_{p}$ is the mass of proton)
and $\Delta \alpha_s$ is the differential polarizability of the clock states.

As given in Table \ref{bbr_shift}, $\Delta \alpha_s$ are obtained as 
$0.1418$ a.u. for the $5f_{5/2} \rightarrow 6p_{1/2}$ transition in 
Cf$^{17+}$ and $-0.1595$ a.u. and $-0.1545$ a.u. for the $6p_{1/2} \rightarrow 5f_{5/2}$ 
transition in Cm$^{15+}$ and Bk$^{16+}$, respectively.
For $5f_{5/2} \rightarrow 5f_{7/2}$ transition, $\Delta \alpha_s$ is significantly
smaller, $-0.0045$ a.u. and $-0.0005$ a.u., respectively in Cm$^{15+}$ and Bk$^{16+}$.
It should be noted that, our computed $\Delta \alpha_s$ for $5f_{5/2} \rightarrow 6p_{1/2}$
transition in Cf$^{17+}$ is of opposite sign compared to that in 
Ref. \cite{porsev-20} due to a larger $\alpha_s$ for $6p_{1/2}$ state in our work.  
Our computed $\alpha_s$ for $5f_{5/2}$ state is, however, in close agreement
with Ref. \cite{porsev-20}. Since $\Delta \alpha_s$ is positive for this 
clock transition, a magic trap drive frequency does not exist for Cf$^{17+}$ ion.

Further, we evaluated $\Omega$ for the proposed clock transitions in 
$\rm Cm^{15+}$ and $\rm Bk^{16+}$. To the best of
our knowledge, the present work provides the first theoretical estimates of $\Omega$
for Cm$^{15+}$ and Bk$^{16+}$ ions. For $6p_{1/2} \rightarrow 5f_{5/2}$ transition,
$\Omega$ is obtained to be 117.64 MHz and 67.94 MHz for Cm$^{15+}$ and Bk$^{16+}$,
respectively. In contrast, substantially larger values, 425.21 MHz and 1391.28 MHz,
respectively are obtained for $5f_{5/2} \rightarrow 5f_{7/2}$ transition in these ions.
This considerably larger $\Omega$ could be attributed to their extremely
small $\Delta \alpha_s$. Although these small $\Delta \alpha_s$ reflect a strong
cancellation of AC Stark contribution
arising from trapping field, they simultaneously require much higher RF drive
frequencies to achieve a complete cancellation of the micromotion-induced shift.
So, the present analysis shows that the micromotion-induced frequency shift
can be effectively suppressed in the proposed clock transitions in 
Cm$^{15+}$ and Bk$^{16+}$ through operation at corresponding 
magic trap drive frequencies.

The magnetic-field induced frequency shifts are important sources of systematic
uncertainty in the optical atomic clocks. In the weak-field regime, the Zeeman
splitting to energy of a state is given by $\Delta E_M=C_{M1}B+C_{M2}B^2$, 
where $C_{M1}=\mu_Bg_JM$ ($\mu_B$ is the 
Bohr magneton and $g_J$ is the Lande g-factor) is the
first-order Zeeman coefficient. The first-order Zeeman shift can be suppressed 
by choosing clock transitions with $M_F=0$ states or by averaging it with 
equal and opposite first-order shifts. The quadratic Zeeman shift then leads to 
the dominant frequency shift. The coefficient $C_{M2}$ can be evaluated, 
within second-order perturbation theory \cite{sobelman-79}, as
\begin{equation}
C_{M2}(F,M) = -(\mu_B g_J)^2 \sum_{F'\neq F}
\frac{\left|\langle F'M|J_z|FM\rangle\right|^2}{E_{F'}-E_F},
\end{equation}
where only levels with $F'=F\pm1$ contribute. The expressions used in our 
calculations and the calculated hyperfine splitting energies and $C_{M2}$ 
coefficients are provided in the SM.

For $^{251}\mathrm{Cf}^{17+}$, since the nuclear spin ($I$) is $1/2$, the hyperfine
structure is determined solely by the magnetic-dipole HFS constant $A$.
Using our calculated $A/g_I$ values (Table \ref{mhfs_table}) and the nuclear
magnetic moment $\mu=-0.571\,\mu_N$ \cite{weber-23}, the smallest differential 
coefficient is obtained to be $\Delta C_{M2}=-5.87 \; \mathrm{kHz}/(\mathrm{mT})^2$
for $5f_{5/2} \rightarrow 6p_{1/2}$ with $F=0 \rightarrow F=2$ or $F=1 \rightarrow F=3$ 
clock transitions.
For a clock frequency of $4.819\times10^{14}$ Hz and the magnetic field of
$10 \;\mu\mathrm{T}$, this corresponds to a fractional frequency shift (FFS) of
$-1.2173\times10^{-15}$. Our computed FFS in good agreement with 
Ref. \cite{porsev-20}.

For $^{247}\mathrm{Cm}^{15+}$ and $^{249}\mathrm{Bk}^{16+}$, since the data on 
the nuclear quadrupole moments are not available, to evaluate the hyperfine 
splitting energies we use the dominant magnetic-dipole interaction. We used the nuclear 
magnetic moments $\mu=0.36 \;\mu_N$ \cite{stone-05} and $2.0\;\mu_N$ \cite{stone-05},
respectively, to obtain the smallest $\Delta C_{M2}$ for $6p_{1/2} \rightarrow 5f_{5/2}$ 
and $5f_{5/2} \rightarrow 5f_{7/2}$ transitions. For Cm$^{15+}$,
we get $\Delta C_{M2}$ as $103.575$ and $255.65 \; \mathrm{kHz}/(\mathrm{mT})^2$, with
corresponding fractional shifts obtained as $7.960\times10^{-15}$ and $5.076\times10^{-14}$ 
at $1.301\times10^{15}$ and $5.037\times10^{14}$ Hz clock frequencies, respectively.
For Bk$^{16+}$, the corresponding $\Delta C_{M2}$s are $21.151$ and
$32.655 \; \mathrm{kHz}/(\mathrm{mT})^2$, leading to fractional shifts of
$4.945\times10^{-15}$ and $5.711\times10^{-15}$ at clock frequencies of
$4.277\times10^{14}$ and $5.718\times10^{14}$ Hz, respectively. These results may 
be considered as estimates because the electric-quadrupole correction may 
modify the hyperfine splittings and, hence, $\Delta C_{M2}$. More precise 
predictions will require improved knowledge of the magnetic-field strength 
and nuclear magnetic moments.

{\em Sensitivity to variation in the fine-structure constant--}We now present and
discuss our results on sensitivity coefficient ($q$) and enhancement
factor ($K_{\alpha}$) for the proposed clock transitions.
The sensitivity coefficient is extracted by computing the gradient
\begin{equation}
	q = \frac{d \omega}{dx}\Big|_{x= 0} ,
\label{omega}
\end{equation}
where  $\omega$ is the atomic transition frequency
and $x = \left(\frac{\alpha}{\alpha_0}\right)^2 - 1 $ is a perturbation
arising due to change in $\alpha$ from its current value $\alpha_0$.
The enhancement factor can be obtained from the sensitivity coefficient
using the relation $K_{\alpha} = 2q/\omega_0$.

\begin{table}
	\caption{Relativistic sensitivity coefficients ($q$) (in cm$^{-1}$) and
	enhancement factors ($K_{\alpha}$) for the proposed clock transitions
	in $\rm Cf^{17+}$, $\rm Cm^{15+}$ and $\rm Bk^{16+}$ HCIs with respect
	to their ground state.
	The uncertainties quoted for $q^{\rm CCSD}$ correspond to the asymptotic
	standard errors of the fit.}
  \begin{ruledtabular}
  \begin{tabular}{cccccc}
	  States & $q^{\rm CCSD}$ & Breit+QED & $q^{\rm final}$ & $K_{\alpha}$ & Others \\
      \hline
      \multicolumn{6}{c}{$\rm Cf^{17+}$} \\
      $5f_{5/2}$ &   0             &   0    &     0     &   0   &   0       \\
      $6p_{1/2}$ &  $-449718(743)$ & $-854$ & $-450572$ & $-55.3$ &  $-449750^{\rm a}$  \\
      $6p_{3/2}$ &  $-115083(11)$  & $-103$ & $-115186$ &  $-1$   &  $-115650^{\rm a}$  \\
      $5f_{7/2}$ &   17667(19)     &   0    &   17667   &  1.6    &
      $17900^{\rm a}$   \\
      \hline
      \multicolumn{6}{c}{$\rm Cm^{15+}$} \\
      $6p_{1/2}$ &     0         &   0   &   0      &    0  &   0     \\
      $5f_{5/2}$ &  363891(1142) &  692  &  364583  &  16.8 & $367000^{\rm b}$  \\
      $5f_{7/2}$ &  377463(1063) &  642  &  378105  &  12.6 & $380000^{\rm b}$  \\
      $6p_{3/2}$ &  267941(12608)&  536  &  268477  &   2.9 &          \\
      \hline
      \multicolumn{6}{c}{$\rm Bk^{16+}$} \\
      $6p_{1/2}$ &        0       &   0   &     0   &  0    &  0    \\
      $5f_{5/2}$ &   405387(1207) &  770  &  406157 & 57.5  & $403000^{\rm b}$ \\
      $5f_{7/2}$ &   420950(1125) &  674  &  421624 & 25.4  & $419000^{\rm b}$  \\
      $6p_{3/2}$ &   300037(866)  &  840  &  300877 &  2.9  &          \\
        \end{tabular}
        \end{ruledtabular}
        \begin{tabbing}
$^{\rm a}$Ref.\cite{berengut-12}-CI + MBPT,
$^{\rm b}$Ref.\cite{dzuba-24}-LCCSD. \\
\end{tabbing}
\label{q_k}
\end{table}

To extract $q$, transition energies were calculated for several values
of $x$ in the vicinity of $x=0$ and $q$ is extracted from fitting
the $E$ vs. $x$ data. In all HCIs, our calculated energy exhibit an 
excellent linearity (Fig. \ref{linear_q} in SM), demonstrating that 
higher-order terms in the expansion of $\omega$ are negligible over 
the investigated range. And hence, $q$ can be reliably identified as 
the slope of $E$ vs. $x$ curve.
Since the fitting procedure utilizes multiple calculated points, it is
less susceptible to numerical fluctuations than the two-points 
finite-difference estimates.
Among the higher-order corrections in $q$ (discussed in SM), 
VP has the largest contribution, about 0.3\% for $6p_{3/2}$ 
state in $\rm Cm^{15+}$.
Although small, these are important for obtaining reliable $q$ 
at accuracy commensurate with precision frequency standards.

For the HCIs considered here, a very few theoretical results on $q$ are
available in the literature. Our computed $q$ is consistent with previously
reported values for all HCIs, with minor differences due to more accurate
treatment of electron correlation in our work. For example, in the case 
of $\rm Cf^{17+}$, our computed $q$ differ by $\approx$
0.2\%, 0.4\%, and $-1.3$\% for corresponding states from CI+MBPT 
calculation \cite{berengut-12}. 
Similarly, for $\rm Cm^{15+}$, our calculated $q$ for
$5f_{5/2}$ and $5f_{7/2}$ states differ by $\approx$ $-0.7$\% and
$-0.5$\%, respectively, from LCCSD calculation \cite{dzuba-24}. Whereas,
for $\rm Bk^{16+}$, the corresponding differences are 0.8\%
and 0.6\%, respectively. These deviations could be attributed to the
inclusion of non-linear CC terms in present calculations, which
are absent in the LCCSD method adopted in Ref. \cite{dzuba-24}.

Our reported $K_{\alpha}$ further highlights the suitability of the proposed
clock transitions for probing possible variation in $\alpha$.
Transitions with larger $|K_{\alpha}|$ provide stronger experimental
signature for probing variation.
As evident from Table \ref{q_k}, in particular, the established 
$5f_{5/2} \rightarrow 6p_{1/2}$ clock transition in $\rm Cf^{17+}$ 
exhibits a remarkably large $K_{\alpha}$ of $-55.3$, making it one of t
he most sensitive optical clock transitions proposed to date for 
the test of $\alpha$-variation.
Our obtained $K_{\alpha}$ is in good agreement with the value, $-48.1$, 
reported in Ref. \cite{berengut-12}.
For $\rm Cm^{15+}$, $6p_{1/2} \rightarrow5f_{5/2}$ transition possesses 
a $K_{\alpha}$ of 16.8, whereas the $5f_{5/2}\rightarrow5f_{7/2}$ has 
a comparatively smaller value, 1.61.
This is in good agreement with the values, 15 and 2, reported 
in Ref. \cite{dzuba-24}, respectively.
In the case of $\rm Bk^{16+}$, our $K_{\alpha}$ associated 
with $6p_{1/2}\rightarrow5f_{5/2}$ transition increases significantly 
to 57.5, while $5f_{5/2}\rightarrow5f_{7/2}$ transition exhibits a value 
of 1.62. The $K_{\alpha}$ obtained in present work are substantially 
larger than those currently used in the state-of-the-art optical-clock 
comparisons for probing the variation in $\alpha$.
For example, the comparison between E2 and E3 transitions in $^{171}$Yb$^{+}$,
one of the most sensitive laboratory probes of $\alpha$ variation \cite{lange-21},
is characterized by $|\Delta K_{\alpha}| \approx 7$.
In contrast, $5f_{5/2}\rightarrow6p_{1/2}$ transition in $\rm Cf^{17+}$ 
and $6p_{1/2}\rightarrow5f_{5/2}$ transition in $\rm Bk^{16+}$ exhibit
$|K_{\alpha}|$ exceeding 50. Such large values highlight the exceptional potential
of these HCIs for future optical-clock experiments aimed at probing possible
temporal variation of fundamental constants.


{\em Conclusion--}We have carried out a comprehensive FSRCC investigation 
of the spectroscopic properties, systematic shifts, and sensitivity to $\alpha$ 
variation in $\rm Cf^{17+}$, $\rm Cm^{15+}$, and $\rm Bk^{16+}$ HCIs. 
To improve the accuracy of computed properties further, corrections from 
Breit interaction, QED effects and perturbative triples are incorporated.
Further, we have estimated the systematic frequency shifts for the proposed 
clock transitions.
Our obtained strongly suppressed frequency shifts indicate that the proposed 
clock transitions exhibit a remarkable robustness against external perturbations. 
These characteristics satisfy a key requirement for the realization of 
ultra-high-accuracy optical clocks based on HCIs.

The relativistic sensitivity coefficients and enhancement factors determined 
in our work further reveal an exceptional sensitivity of considered HCIs to 
possible $\alpha$ variation. In particular, $5f_{5/2} \rightarrow 6p_{1/2}$ 
clock transition in $\rm Cf^{17+}$ exhibits an enhancement factor of 
$K_{\alpha}=-55.3$, placing it among the most sensitive optical-clock 
transitions proposed to date. Similarly, $6p_{1/2} \rightarrow 5f_{5/2}$ 
transition in $\rm Bk^{16+}$ possesses an enhancement factor of $K_{\alpha}=57.5$, 
demonstrating a comparable potential. 
Further, a comparison of the two proposed clock transitions in $\rm Cm^{15+}$ and $\rm 
Bk^{16+}$ reveals a clear distinction in their optimal applications. The 
$6p_{1/2} \rightarrow 5f_{5/2}$ transition is observed to exhibit a giant 
enhancement factor and is therefore particularly attractive for laboratory 
searches for the variation in $\alpha$. In contrast, $5f_{5/2} \rightarrow 5f_{7/2}$ 
transition displays reduced systematic shifts, making it a more favorable 
transition for the realization of an ultra-precise optical
frequency standard. The upper bound to theoretical uncertainty in E1, E2, 
M1 and HFS matrix elements is 1\%, whereas it is 2\% in the case 
of $\alpha_s$. 


{\em Acknowledgments--}We thank Suraj Pandey for useful
discussions. PG acknowledges funding support from UGC (BININ04154142), 
Govt. of India. BKM, acknowledges the funding support from SERB, 
DST (Grant No. CRG/2022/003845). 
Calculations are performed using High Performance Computing cluster 
Tejas at Indian Institute of Technology Delhi and PARAM Rudra 
at Inter-University Accelerator Centre, New Delhi.


\bibliography{hci}

\end{document}


\title{Supplemental Material: Fock-space relativistic coupled-cluster study of systematic shifts and
       fine-structure constant sensitivity in {\rm Cf$^{17+}$}, {\rm Cm$^{15+}$}
       and {\rm Bk$^{16+}$} highly charged ions}

\author{Palki Gakkhar}
\affiliation{Department of Physics, Indian Institute of Technology,
             Hauz Khas, New Delhi 110016, India}


\author{D. Angom}
\affiliation{Department of Physics, Manipur University,
             Canchipur 795003, Manipur, India}

\author{B. K. Mani}
\email{bkmani@physics.iitd.ac.in}
\affiliation{Department of Physics, Indian Institute of Technology,
             Hauz Khas, New Delhi 110016, India}

\begin{abstract}

\end{abstract}

\maketitle

\section{Methodology}

We have employed a FSRCC theory to calculate the
many-body wavefunctions which are used to calculate the relevant clock's
properties and systematic shifts.
For a one-valence atomic system, the many-body wavefunction is obtained
by solving the eigen value equation
\begin{equation}
  H^{\rm DC}|\Psi_v \rangle = E_v |\Psi_v \rangle,
  \label{hdc_eqn}
\end{equation}
where $|\Psi_v \rangle$ is the exact atomic state of a one-valence system
and $E_v$ is the corresponding energy. The Hamiltonian $H^{\rm DC}$ is
the Dirac-Coulomb Hamiltonian \cite{sucher-80}.
In FSRCC, $|\Psi_v \rangle$ is expressed as
\begin{equation}
 |\Psi_v\rangle = e^{T^{(0)}} \left [  1 + S^{(0)} \right ] |\Phi_v\rangle,
 \label{psi_unptrb}
\end{equation}
where $T^{(0)}$ and $S^{(0)}$ represent the unperturbed coupled-cluster (CC)
operators for closed-shell and one-valence sectors, respectively.
These are obtained by solving the nonlinear CC equations discussed
in our previous works: Refs. \cite{mani-09, mani-17} for closed-shell
and Ref. \cite{mani-10, mani-17} for one-valence systems. $|\Phi_v\rangle$ is
the Dirac-Fock reference state for one-valence systems and is obtained by
adding an electron to the closed-shell reference state,
as $|\Phi_v \rangle = a^\dagger_v|\Phi_0\rangle$.
We have used CCSD approximation where $T^{(0)}$ and $S^{(0)}$ CC operators
are truncated to have only $one-$ and $two-body$ operators as these alone
provide a good description of the atomic properties. However, to improve
the accuracy of our calculations further, we include the contributions
from triples excitations perturbatively discussed in our 
previous works \cite{mani-09, mani-10}.

\section{Basis set and convergence of properties}

To get reliable properties results using FSRCC theory, it is essential 
to use accurate single-electron basis. 
In this work, we have employed Gaussian-type
orbitals (GTOs) \cite{mohanty-91} as the single-electron basis. GTOs
form a finite basis set where single-particle wavefunctions are
expressed as linear combinations
of Gaussian-type functions (GTFs). The large component of
the wavefunction is expressed as
\begin{equation}
  g^L_{\kappa p} (r) = C^L_{\kappa i} r^{n_\kappa} e^{-\alpha_p r^2},
\end{equation}
where $p = 0$, $1$, $2$, $\ldots$, $N$ refers to the GTO index and
$N$ is the total number of GTFs. The exponent $\alpha_p$ is further
expressed as $\alpha_0 \beta^{p-1}$, where $\alpha_{0}$ and $\beta$
are two independent parameters. Parameters $\alpha_{0}$ and $\beta$ 
are optimized to reproduce the self-consistent field (SCF)
and single-particle energies obtained from B-spline \cite{oleg-16} basis. 
Table \ref{basis} lists the optimized Gaussian parameters and \ref{conv-basis} 
compares GTOs single-particle and SCF energies with B-spline data. As evident
from the table, GTOs energies are in good agreement with B-spline energies. This
confirms that a good quality basis is used in our FSRCC calculations.

\begin{figure*}
\includegraphics[scale = 0.21, angle=-90]{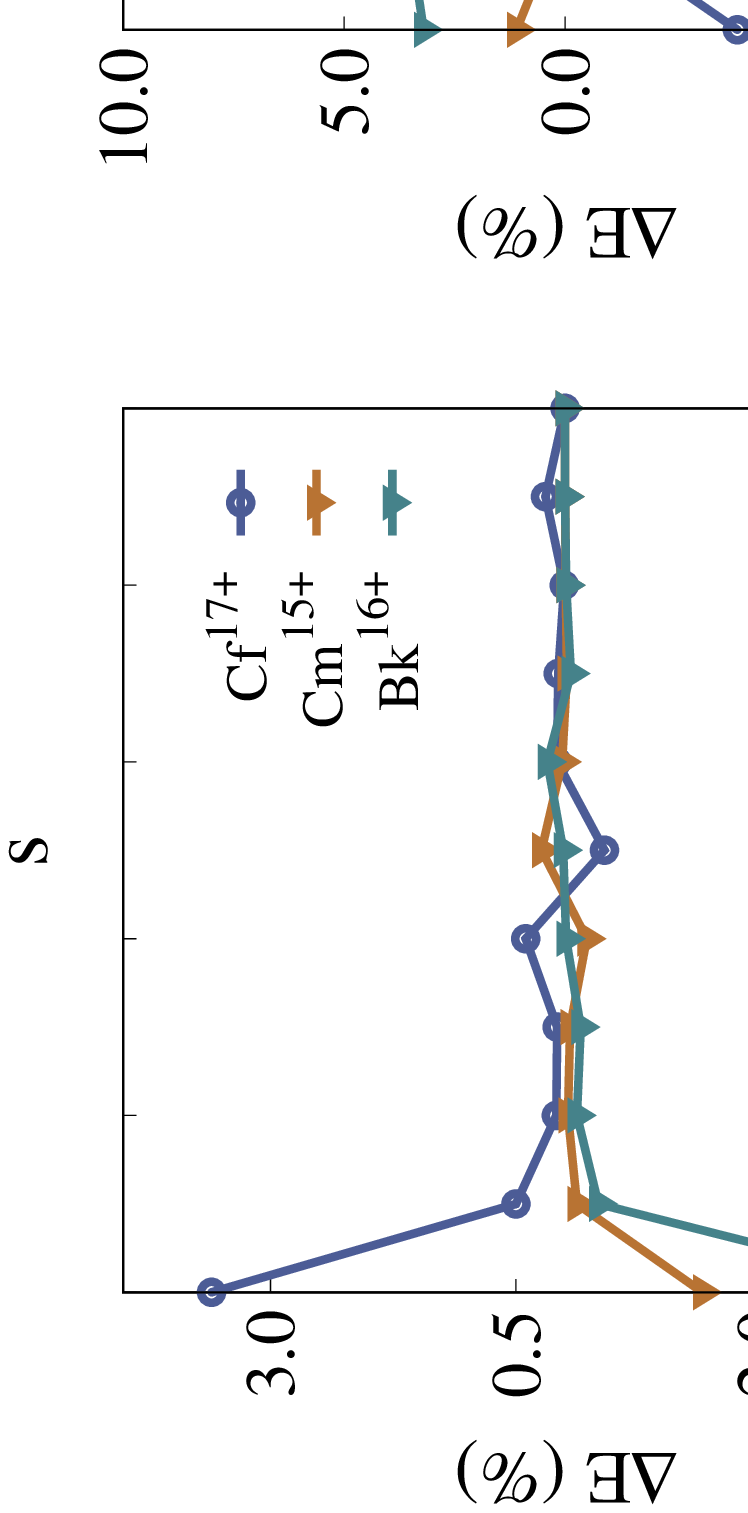}
\caption{Panels illustrate the basis-set convergence for excitation 
	energies of first excited states, $6p_{1/2}$ in $\rm Cf^{17+}$ and 
	$5f_{5/2}$ in $\rm Cm^{15+}$ and $\rm Bk^{16+}$ HCIs. Orbitals in 
	$s$, $p$, $d$, $f$, $g$, $h$, $i$, and $j$ symmetries are augmented 
	until the excitation energy converges. 
}
\label{conv-diag}
\end{figure*}

Since GTO forms a mathematically incomplete basis, it is essential to check 
the convergence of the computed properties with basis size. 
Fig. \ref{conv-diag} shows the convergence of excitation energy with
augmentation of orbitals in $s$, $p$, $d$, $f$, $g$, $h$, $i$, and $j$ 
symmetries. As we observed from the figure, our excitation energy is well 
converged with respect to orbitals from each symmetries and the final 
converged basis contains $22s22p18d18f17g14h22i22j$ orbitals. However, from 
our calculations, we find that properties converge even faster with a basis
set having $14s13p11d10f12g8h8i8j$ orbitals.
This finalized basis is then used for the calculation of other properties. 
Corrections arising from the Breit interaction, QED effects and perturbative 
triple excitations are also evaluated and incorporated into the final results 
to improve their accuracy.


\begin{table*}
   \caption{The $\alpha_0$ and $\beta$ parameters of the even tempered GTO basis 
    used in our calculations.}
   \label{basis}
   \begin{ruledtabular}
   \begin{tabular}{ccccccccc}
     Atom & \multicolumn{2}{c}{$s$} & \multicolumn{2}{c}{$p$} & 
     \multicolumn{2}{c}{$d$} & \multicolumn{2}{c}{$f$} \\
     \cline{2-3} \cline{4-5} \cline{6-7} \cline{8-9}
     & $\alpha_{0}$  & $\beta$ & $\alpha_{0}$ & $\beta$  
     & $\alpha_{0}$  & $\beta$  & $\alpha_{0}$  & $\beta$ \\
    \hline
     ${\rm Cf^{17+}}$  &\, 0.00400 &\, 1.825 &\,0.00428 &\, 1.812 &\, 0.00555 &\, 1.875 &\, 0.00435 &\,1.885 \\
     ${\rm Cm^{15+}}$  &\, 0.00340 &\, 1.783 &\,0.00440 &\, 1.823 &\, 0.00555 &\, 1.875 &\, 0.00435 &\,1.885 \\
     ${\rm Bk^{16+}}$  &\, 0.00300 &\, 1.825 &\,0.00438 &\, 1.855 &\, 0.00355 &\, 1.875 &\, 0.00335 &\,1.885 \\
   \end{tabular}
   \end{ruledtabular}
\end{table*}

\begin{table*}
\begin{center}
\caption{Comparison of single-particle and SCF energies (in Hartree) from GTO with
         B-Spline results. }
\begin{ruledtabular}
\begin{tabular}{ccccccc}
  Orbitals  & \multicolumn{2}{c}{\rm Cf$^{17+}$}  & \multicolumn{2}{c}{\rm Cm$^{15+}$} 
  & \multicolumn{2}{c}{\rm Bk$^{16+}$} \\
  \cline{2-3} \cline{4-5} \cline{6-7}
  & GTO   & B-spline  & GTO & B-spline & GTO & B-spline \\
 \hline
$1s_{1/2}$  & 5011.01288  & 5011.03256 & 4759.46911 & 4759.40577 & 4883.95074 & 4884.00019 \\
$2s_{1/2}$  &  977.86177  &  977.86519 &  920.08034 &  920.06781 &  948.60129 &  948.60991 \\
$2p_{1/2}$  &  944.60465  &  944.61029 &  887.98170 &  887.98441 &  915.91749 &  915.92779 \\ 
$2p_{3/2}$  &  750.91545  &  750.91510 &  713.64752 &  713.64746 &  732.14492 &  732.14424 \\      
$3s_{1/2}$  &  265.60734  &  265.60811 &  247.24488 &  247.24190 &  256.30067 &  256.30269 \\ 
$3p_{1/2}$  &  250.41982  &  250.42134 &  232.62204 &  232.62284 &  241.39478 &  241.39757 \\
$3p_{3/2}$  &  204.62434  &  204.62420 &  191.60805 &  191.60809 &  198.05151 &  198.05127 \\
$3d_{3/2}$  &  181.10269  &  181.10257 &  168.83682 &  168.83686 &  174.90626 &  174.90604 \\
$3d_{5/2}$  &  172.23562  &  172.23551 &  160.78358 &  160.78362 &  166.45368 &  166.45347 \\ 
$4s_{1/2}$  &   82.07445  &   82.07466 &   74.63659 &   74.63576 &   78.29937 &   78.29990 \\
$4p_{1/2}$  &   75.03788  &   75.03831 &   67.90586 &   67.90608 &   71.41601 &   71.41676 \\ 	       
$4p_{3/2}$  &   62.52833  &   62.52828 &   56.81944 &   56.81944 &   59.63760 &   59.63751 \\ 
$4d_{3/2}$  &   51.44848  &   51.44844 &   46.17107 &   46.17107 &   48.77410 &   48.77402 \\
$4d_{5/2}$  &   49.18689  &   49.18685 &   44.14779 &   44.14779 &   46.63425 &   46.63418 \\
$4f_{5/2}$  &   34.56576  &   34.56571 &   30.09211 &   30.09209 &   32.29597 &   32.29589 \\
$4f_{7/2}$ &   33.95263  &   33.95259 &   29.55219 &   29.55218 &   31.72030 &   31.72023 \\
$5s_{1/2}$  &   29.52684  &   29.52689 &   25.83993 &   25.83970 &   27.65303 &   27.65317 \\
$5p_{1/2}$  &   26.49859  &   26.49871 &   22.99086 &   22.99102 &   24.71475 &   24.71494 \\
$5p_{3/2}$  &   23.16480  &   23.16481 &   20.12674 &   20.12675 &   21.62302 &   21.62299 \\
$5d_{3/2}$  &   18.49807  &   18.49809 &   15.73861 &   15.73864 &   17.09570 &   17.09569 \\
$5d_{5/2}$  &   17.95858  &   17.95859 &   15.27849 &   15.27852 &   16.59694 &   16.59692 \\
$6s_{1/2}$  &   13.57324  &   13.57327 &   11.46675 &   11.46668 &   12.50055 &   12.50059 \\

\hline
Virtuals & GTO   & B-spline  & GTO & B-spline & GTO & B-spline \\
\hline
$6p_{1/2}$  &   11.80981  &   11.83081 &   9.84562 &   9.86589 &   10.80834 &   10.82901 \\
$6p_{3/2}$  &   10.78452  &   10.79915 &   9.01054 &   9.02490 &    9.88144 &    9.89593 \\
$5f_{5/2}$  &   11.84832  &   11.91451 &   9.61207 &   9.67432 &   10.70736 &   10.77163 \\
$5f_{7/2}$ &   11.75515  &   11.81812 &   9.54026 &   9.59955 &   10.62519 &   10.68638 \\
\hline
	$E_{\rm SCF}$ & -32963.45287 & -32963.52204 & -31253.75038 & -31253.78154 & -32100.84541 
    & -32100.97833 \\
 \end{tabular}
\end{ruledtabular}
\label{conv-basis}
\end{center}
\end{table*}    

\section{Breit, QED and Perturbative Triples Contributions to 
Excitation energy, Lifetime and HFS constants}

\begin{figure*}
\includegraphics[scale = 0.42, angle=-90]{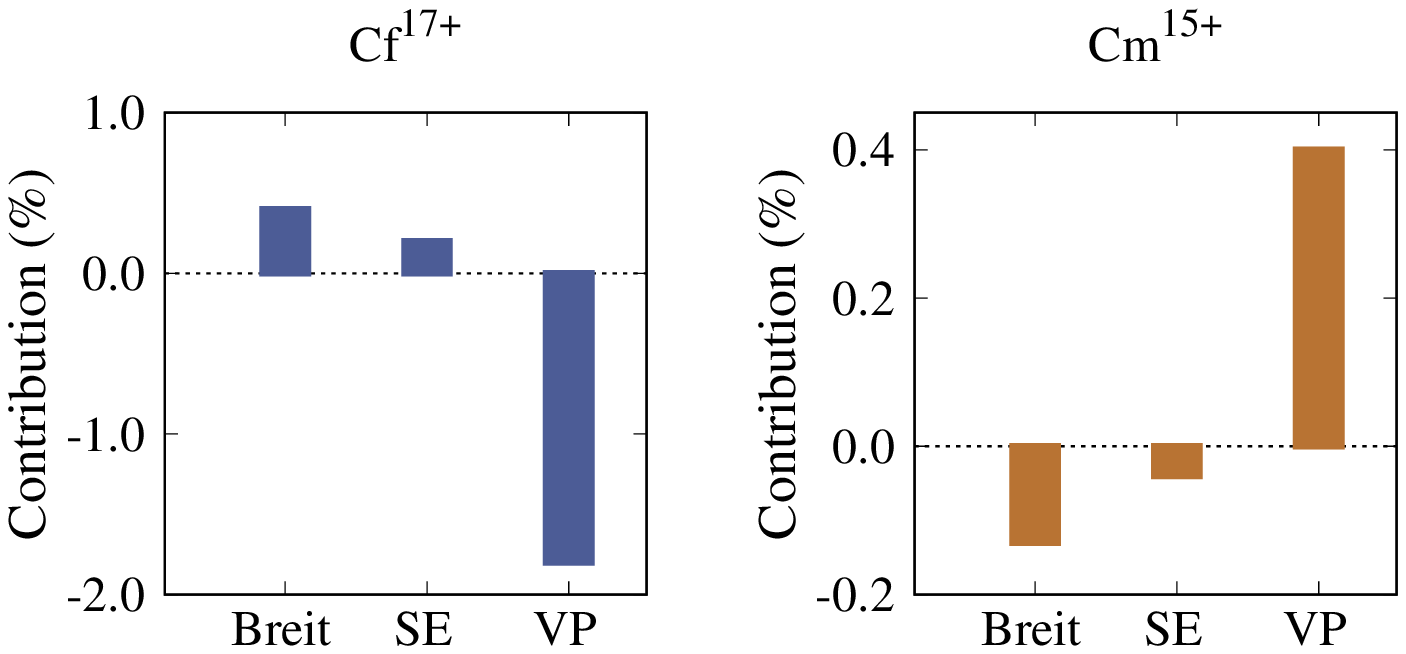}
\caption{Dominant contributions arising from Breit interaction, self-energy, and 
vacuum-polarization effects to the excitation energies of low-lying states of
	$\rm Cf^{17+}$, $\rm Cm^{15+}$, and $\rm Bk^{16+}$ respectively.}
\label{energy_bqed}
\end{figure*}

To further improve the accuracy of the calculated properties, corrections
arising from the Breit interaction and QED effects, such as vacuum polarization 
and self-energy are incorporated. In this section, we provide some quantitative 
assessment of their contributions to these properties. 

The dominant contributions from these higher-order relativistic effects 
to excitation energy are shown in Fig. \ref{energy_bqed}. Among the low-lying
states, the largest Breit and QED corrections are observed for the first excited states.
The Breit interaction contributes up to 0.4\%, $-0.1$\%, and $-0.3$\% to the first excited
states of $\rm Cf^{17+}$, $\rm Cm^{15+}$, and $\rm Bk^{16+}$, respectively. 
Whereas, the corresponding QED corrections amount to be $-1.6$\%, 0.4\%, and 1.2\%. 
Although these corrections are considerably smaller than the dominant 
residual Coulomb interaction, their inclusion is important for achieving 
high-precision spectroscopic predictions.

\begin{figure*}
\includegraphics[scale = 0.42, angle=-90]{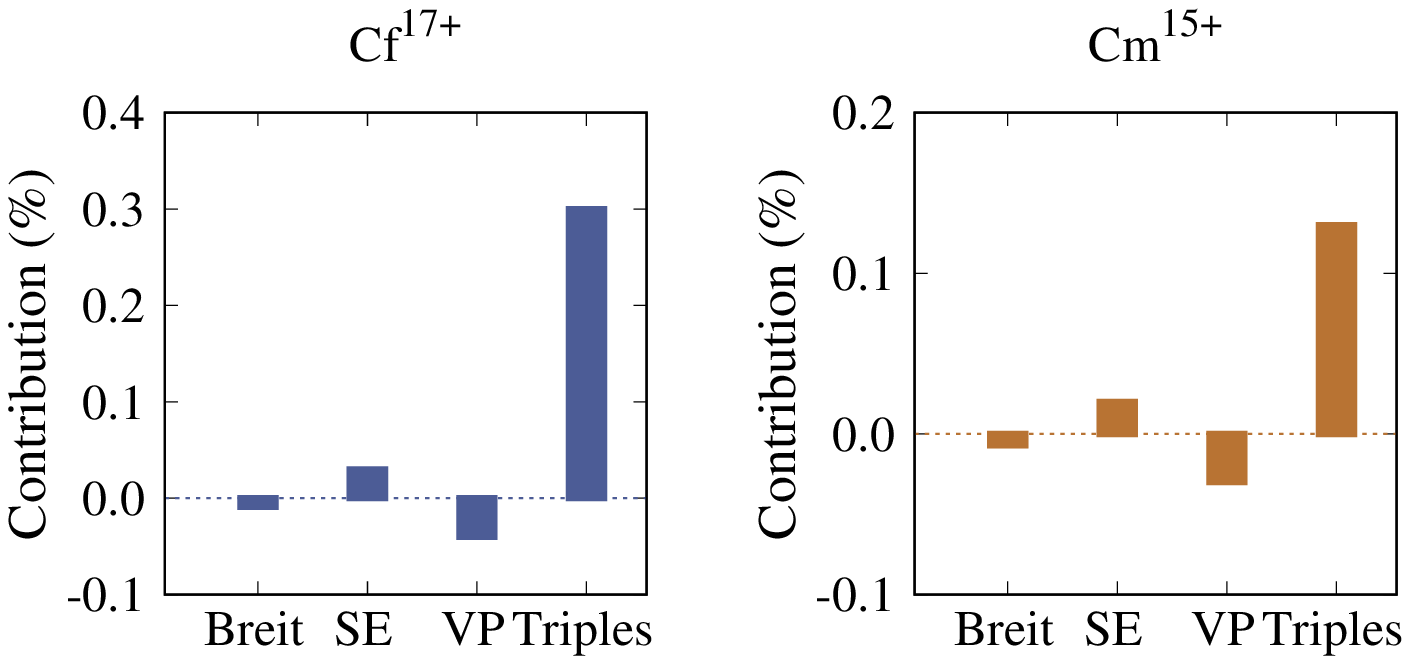}
\caption{Dominant contributions arising from Breit interaction, self-energy, and 
vacuum-polarization effects to the lifetime of low-lying states of
	$\rm Cf^{17+}$, $\rm Cm^{15+}$, and $\rm Bk^{16+}$ respectively.}
\label{lifetime_bqed}
\end{figure*}

Fig \ref{lifetime_bqed} shows the corresponding dominant contributions for lifetime. 
As it is discernible from the figure, contributions from these corrections are 
generally negligible for $6p_{3/2}$ and $5f_{7/2}$ states 
in all three HCIs. The largest Breit contribution is found to be $-0.009\%$ for
$6p_{1/2}$ state of $\rm Cf^{17+}$.  Similarly, the largest QED contribution is 
approximately $-0.01\%$ for the first excited states, $6p_{1/2}$ in $\rm Cf^{17+}$ 
and $5f_{5/2}$ in both $\rm Cm^{15+}$ and $\rm Bk^{16+}$. Among all the considered states, 
the largest perturbative triples correction to the lifetime is found for
$6p_{1/2}$ state of $\rm Cf^{17+}$, amounting to approximately 0.3\%.

\begin{figure*}
\includegraphics[scale = 0.42, angle=-90]{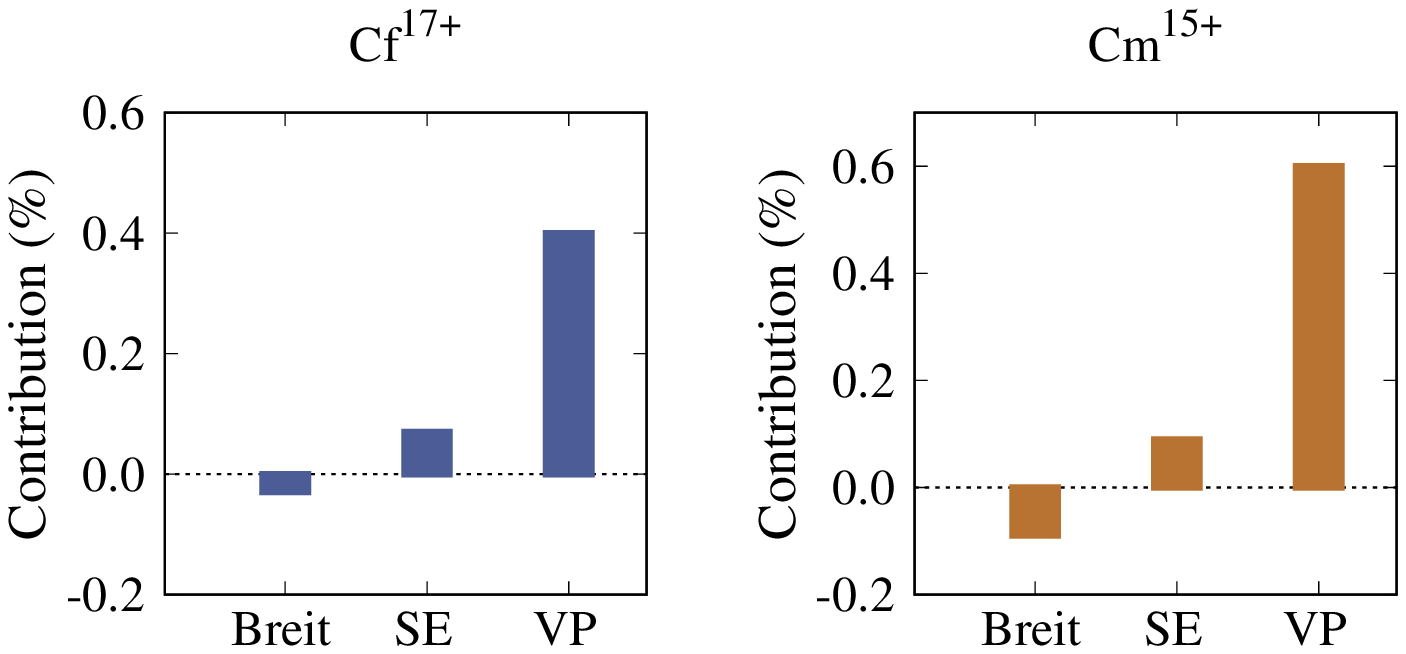}
\caption{Largest contributions arising from the Breit interaction, self-energy and 
vacuum-polarization effects to the magnetic dipole hyperfine structure constant in $
\rm Cf^{17+}$, $\rm Cm^{15+}$ and $\rm Bk^{16+}$ respectively.}
\label{mhfs_bqed}
\end{figure*}

The dominant contributions from Breit interaction, self-energy correction, 
and vacuum polarization to magnetic dipole HFS constants are shown in 
Fig. \ref{mhfs_bqed}. As can be observed from the figure, contributions 
from Breit and self-energy are negligible. In contrast, the vacuum-polarization 
correction exhibits a noticeable contribution of approximately 0.4\% to 
$6p_{1/2}$ in all three systems. This behavior can be attributed to the 
significant electron density of $6p_{1/2}$ state in nuclear region, which 
enhances its sensitivity to short-range QED effects.

\section{Electric dipole polarizability}

In Table \ref{static-pol}, we provide our results on static electric 
dipole polarizability ($\alpha_s$) for clock relevant states using FSRCC theory. The
corrections from Breit, QED and perturbative triples are listed separately
for quantitative assessment of correlation effects. We have also provided 
the dominant contributing E1 transitions contributing to $\alpha_s$. 

\begin{table*}
\begin{center}
\caption{Electric dipole polarizability (a.u.) for $6p_{1/2}$, $5f_{5/2}$ 
	and $5f_{7/2}$ states of {\rm Cf$^{17+}$}, {\rm Cm$^{15+}$} 
	and {\rm Bk$^{16+}$} using sum-over-states approach. The dominant 
	contributing E1 matrix elements are also provided to quantify 
	the nature of electron correlations.}
\begin{ruledtabular}
\label{static-pol}
\begin{tabular}{cccccccccc}
  \multicolumn{9}{c}{\rm Cf$^{17+}$} \\
  \hline
  \multicolumn{3}{c}{\rm 6p$_{1/2}$}  & \multicolumn{3}{c}{\rm 5f$_{5/2}$}  &
  \multicolumn{3}{c}{\rm 5f$_{7/2}$}  \\
  \cline{1-3} \cline{4-6} \cline{7-9}
  E1 transition  & E1amp & $\alpha_s$  &  E1 transition & E1amp & $\alpha_s$ &
  E1 transition  & E1amp & $\alpha_s$ \\
 \hline
$7s_{1/2} \rightarrow 6p_{1/2}$ & 0.3520 & 0.0099 & $6d_{3/2} \rightarrow 5f_{5/2}$ & $0.7937$ & 0.0247 &
$6d_{5/2} \rightarrow 5f_{7/2}$ & 0.8842 & 0.0225  \\
$8s_{1/2} \rightarrow 6p_{1/2}$ & 0.0238 & 0.0    & $6d_{5/2} \rightarrow 5f_{5/2}$ & $0.1737$ & 0.0011 &
$5g_{7/2} \rightarrow 5f_{7/2}$ & $0.1963$ & 0.0007  \\
$6d_{3/2} \rightarrow 6p_{1/2}$ & 1.2494 & 0.1886 & $5g_{7/2} \rightarrow 5f_{5/2}$ & $1.1016$  & 0.0259 &
$5g_{9/2} \rightarrow 5f_{7/2}$ & $1.2505$ & 0.0300  \\                
\hline
Core               & & 0.8907  &&& 0.8907  &&& 0.8907  \\
CCSD               & & 1.0892  &&& 0.9424  &&& 0.9439  \\
CCSD(T)            & & 1.0870  &&& 0.9443  &&& 0.9458 \\
CCSD(T)+Breit      & & 1.0870  &&& 0.9443  &&& 0.9458 \\
CCSD(T)+Breit+QED  & & 1.0859  &&& 0.9441  &&& 0.9454 \\
Other theory & & 0.919$^{\rm a}$    &&& 0.961$^{\rm a}$    &&&          \\
\hline

  \multicolumn{9}{c}{\rm Cm$^{15+}$} \\
  \hline
  \multicolumn{3}{c}{\rm 6p$_{1/2}$}  & \multicolumn{3}{c}{\rm 5f$_{5/2}$}  &
  \multicolumn{3}{c}{\rm 5f$_{7/2}$}  \\
  \cline{1-3} \cline{4-6} \cline{7-9}
  E1 transition  & E1amp & $\alpha_s$  &  E1 transition & E1amp & $\alpha_s$ &
  E1 transition  & E1amp & $\alpha_s$ \\
 \hline
$7s_{1/2} \rightarrow 6p_{1/2}$ & 0.3779 & 0.0135 & $6d_{3/2} \rightarrow 5f_{5/2}$ & $0.9958$ & 0.0497 &
$6d_{5/2} \rightarrow 5f_{7/2}$ & 1.0668 & 0.0422  \\
$6d_{3/2} \rightarrow 6p_{1/2}$ & 1.3311 & 0.2448 & $6d_{5/2} \rightarrow 5f_{5/2}$ & $0.1913$ & 0.0017 &
$5g_{7/2} \rightarrow 5f_{7/2}$ & $0.2335$ & 0.0012  \\
                                &        &        & $5g_{7/2} \rightarrow 5f_{5/2}$ & $1.2033$  & 0.0416 &
$5g_{9/2} \rightarrow 5f_{7/2}$ & $1.4480$ & 0.0459  \\                
\hline
Core               & & 1.1045  &&& 1.1045  &&& 1.1045  \\
CCSD               & & 1.3628  &&& 1.1975  &&& 1.1938  \\
CCSD(T)            & & 1.3587  &&& 1.1987  &&& 1.1950  \\
CCSD(T)+Breit      & & 1.3587  &&& 1.1987  &&& 1.1950 \\
CCSD(T)+Breit+QED  & & 1.3577  &&& 1.1982  &&& 1.1937 \\
Other theory & & 1.1998$^{\rm b}$    &&& 1.2460$^{\rm b}$    &&& 1.2433$^{\rm b}$         \\
\hline

  \multicolumn{9}{c}{\rm Bk$^{16+}$} \\
  \hline
  \multicolumn{3}{c}{\rm 6p$_{1/2}$}  & \multicolumn{3}{c}{\rm 5f$_{5/2}$}  &
  \multicolumn{3}{c}{\rm 5f$_{7/2}$}  \\
  \cline{1-3} \cline{4-6} \cline{7-9}
  E1 transition  & E1amp & $\alpha_s$  &  E1 transition & E1amp & $\alpha_s$ &
  E1 transition  & E1amp & $\alpha_s$ \\
 \hline
$7s_{1/2} \rightarrow 6p_{1/2}$ & $0.3642$ & 0.0115 & $6d_{3/2} \rightarrow 5f_{5/2}$ & $0.8740$ & 0.0337 &
$6d_{5/2} \rightarrow 5f_{7/2}$ & $0.9583$  & 0.0299  \\
$6d_{3/2} \rightarrow 6p_{1/2}$ & $1.2954$ & 0.2167 & $6d_{5/2} \rightarrow 5f_{5/2}$ & $0.1824$  & 0.0014 &
$5g_{7/2} \rightarrow 5f_{7/2}$ & $0.2144$  & 0.0009 \\
                                &           &        & $5g_{7/2} \rightarrow 5f_{5/2}$ & $1.1084$  & 0.0329 &
$5g_{9/2} \rightarrow 5f_{7/2}$ & $1.3474$  & 0.0371  \\                
\hline
Core               & &  0.9901 &&& 0.9901  &&& 0.9901  \\
CCSD               & &  1.2183 &&& 1.0581  &&& 1.0580 \\
CCSD(T)            & &  1.2146 &&& 1.0592  &&& 1.0591 \\
CCSD(T)+Breit      & &  1.2146 &&& 1.0592  &&& 1.0591 \\
CCSD(T)+Breit+QED  & &  1.2135 &&& 1.0590  &&& 1.0585 \\
Other theory & & 1.0685$^{\rm b}$    &&& 1.1091$^{\rm b}$    &&& 1.1069$^{\rm b}$         \\
 \end{tabular}
\end{ruledtabular}
\begin{tabbing}
$^{\rm a}$Ref.\cite{porsev-20}-CCSDT,
$^{\rm b}$Ref.\cite{dzuba-24}-LCCSD \\
\end{tabbing}

\end{center}
\end{table*}

\section{Zeeman Shift}

The second-order Zeeman shift arising from the coupling 
between hyperfine-structure levels can be evaluated using second-order 
perturbation theory:
\begin{equation}
C_{M2}(F,M) = -(\mu_B g_J)^2 \sum_{F'\neq F} \frac{\left|\langle F'M|J_z 
FM\rangle\right|^2}{E_{F'}-E_F}.
\end{equation}
For the $J_z$ operator, only the terms satisfying $F'=F\pm1$ contribute to the 
second-order Zeeman coefficient. The corresponding matrix elements are expressed 
in terms of angular-momentum coupling coefficients as \cite{sobelman-79}
\begin{equation}
\begin{aligned}
\langle F'M|J_z|FM\rangle = \sqrt{I(I+1)(2I+1)(2F+1)(2F'+1)} \times
\begin{pmatrix}
F & 1 & F' \\
-M & 0 & M
\end{pmatrix}
\begin{Bmatrix}
I & F & J \\
F' & I & 1
\end{Bmatrix}.
\end{aligned}
\end{equation}
The hyperfine energies are calculated as \cite{jonsson_book}
\begin{equation}
E_F^{\rm hfs} = \frac{A}{2}K+\frac{B}{4}N,
\end{equation}
where $A$ and $B$ are the magnetic-dipole and electric-quadrupole hyperfine-structure 
constants, respectively, and
\begin{equation}
K=F(F+1)-I(I+1)-J(J+1),
\end{equation}
\begin{equation}
N= \frac{3K(K+1)-4I(I+1)J(J+1)}{2I(2I-1)J(2J-1)}.
\end{equation}

Using our calculated value of magnetic dipole hyperfine constants ($A$),
we determine the hyperfine-level energies 
$E_F$, which are subsequently employed in the evaluation of the 
second-order Zeeman coefficients. Table \ref{cf_coeff}, \ref{cm_coeff}, 
\ref{bk_coeff} lists the calculated 
hyperfine-level energies and $C_{M2}$ coefficients for the $M=0$ hyperfine levels 
of the proposed clock states of $^{251}\mathrm{Cf}^{17+}$, $^{247}\mathrm{Cm}^{15+}$
and $^{249}\mathrm{Bk}^{16+}$, respectively.

\begin{table}
    \caption{Hyperfine-structure and second-order Zeeman coefficients (C$_{M2}$) for
    $6p_{1/2}$ and $5f_{5/2}$ states with $M=0$ in $\rm ^{251}Cf^{17+}$.
    The values of $\frac{E_F}{h}$ and C$_{M2}$ are given in terms of MHz 
	$\times$ $\rm{g_{I}}$ and $\frac{\rm{kHz} \times \rm{g_I}}{\rm{mT^{2}}}$,
	respectively.}
  \label{cf_coeff}
  \begin{ruledtabular}
  \begin{tabular}{crrcrrcrr}
      \multicolumn{3}{c}{$6p_{1/2}$} & \multicolumn{3}{c}{$5f_{5/2}$}  \\
      \cline{1-3}  \cline{4-6} \cline{7-9}
      $F$ & $\frac{E_F}{h}$  & C$_{M2}$  & $F$ & $\frac{E_F}{h}$  & C$_{M2}$ \\
      \hline
      0  & $-149667.75$ & $-0.109$ & 2  & $-3080$ & $-6.811$  \\
      1  &  49889.25    &   0.109  & 3  &  2200   &   6.811    \\
        \end{tabular}
        \end{ruledtabular}
\end{table}


\begin{table}
    \caption{Hyperfine-structure and second-order Zeeman coefficients (C$_{M2}$) for
    $6p_{1/2}$, $5f_{5/2}$ and $5f_{7/2}$ states with $M=0$ in $\rm ^{247}Cm^{15+}$.
    The values of $\frac{E_F}{h}$ and C$_{M2}$ are given in terms of 
    MHz$\times$ $\rm{g_{I}}$ and $\frac{\rm{kHz} \times \rm{g_I}}{\rm{mT^{2}}}$,
    respectively.}
  \label{cm_coeff}
  \begin{ruledtabular}
  \begin{tabular}{crrcrrcrr}
      \multicolumn{3}{c}{$6p_{1/2}$} & \multicolumn{3}{c}{$5f_{5/2}$}  & \multicolumn{3}{c}{$5f_{7/2}$} \\
      \cline{1-3}  \cline{4-6} \cline{7-9}
      $F$ & $\frac{E_F}{h}$  & C$_{M2}$  & $F$ & $\frac{E_F}{h}$  & C$_{M2}$  & $F$ & $\frac{E_F}{h}$   & C$_{M2}$   \\
      \hline
      4  & $-435363.5$ & $-0.275$ & 2  & $-19263.75$ & $-67.262$ & 1 & $-19500.25$ & $-486.216$ \\
      5  &  356206.5   &   0.275  & 3  & $-15060.75$ &    8.561  & 2 & $-17474.25$ &   139.820  \\
         &             &          & 4  &  $-9456.75$ &   16.209  & 3 & $-14435.25$ &   102.085  \\
         &             &          & 5  &  $-2451.75$ &   15.672  & 4 & $-10383.25$ &    72.863  \\
         &             &          & 6  &    5954.25  &   14.119  & 5 & $-5318.25$  &    55.534  \\
         &             &          & 7  &   15761.25  &   12.699  & 6 &    759.75   &    44.858  \\
         &             &          &    &             &           & 7 &   7850.75   &    37.904  \\
         &             &          &    &             &           & 8 &  15954.75   &    33.151  \\

        \end{tabular}
        \end{ruledtabular}
\end{table}


\begin{table}
    \caption{Hyperfine-structure and second-order Zeeman coefficients (C$_{M2}$) for
    $6p_{1/2}$, $5f_{5/2}$ and $5f_{7/2}$ states with $M=0$ in $\rm ^{249}Bk^{16+}$.
    The values of $\frac{E_F}{h}$ and C$_{M2}$ are given in terms of MHz
    $\times$ $\rm{g_{I}}$ and $\frac{\rm{kHz} \times \rm{g_I}}{\rm{mT^{2}}}$,
    respectively.}
  \label{bk_coeff}
  \begin{ruledtabular}
  \begin{tabular}{crrcrrcrr}
      \multicolumn{3}{c}{$6p_{1/2}$} & \multicolumn{3}{c}{$5f_{5/2}$}  & \multicolumn{3}{c}{$5f_{7/2}$} \\
      \cline{1-3}  \cline{4-6} \cline{7-9}
      $F$ & $\frac{E_F}{h}$  & C$_{M2}$  & $F$ & $\frac{E_F}{h}$  & C$_{M2}$  & $F$ & $\frac{E_F}{h}$   & C$_{M2}$   \\
      \hline
      3  & $-404005.5$ & $-0.030$ & 1  & $-17730$ & $-102.736$ & 0 & $-17640$ & $-1199.359$ \\
      4  &  314226.5   &   0.030  & 2  & $-14578$ &    33.158  & 1 & $-16520$ &    742.462  \\
         &             &          & 3  &  $-9850$ &    24.733  & 2 & $-14280$ &    187.653  \\
         &             &          & 4  &  $-3546$ &    18.279  & 3 & $-10920$ &     95.188  \\
         &             &          & 5  &    4334  &    14.459  & 4 & $-6440$  &     61.563  \\
         &             &          & 6  &   13790  &    12.107  & 5 & $-840$   &     45.396  \\
         &             &          &    &          &            & 6 &   5880   &     36.437  \\
         &             &          &    &          &            & 7 &  13720   &     30.753  \\

        \end{tabular}
        \end{ruledtabular}
\end{table}


\section{Sensitivity Coefficient}

\begin{figure*}
\includegraphics[scale = 0.20, angle=-90]{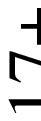}
\caption{Fit of the transition energies vs. $x$ data for $\rm Cf^{17+}$, 
	$\rm Cm^{15+}$, and $\rm Bk^{16+}$ HCIs. Bullet points represent 
	our computed data, whereas linear fit is represented by solid lines.
	}
\label{linear_q}
\end{figure*}

\begin{figure*}
\includegraphics[scale = 0.42, angle=-90]{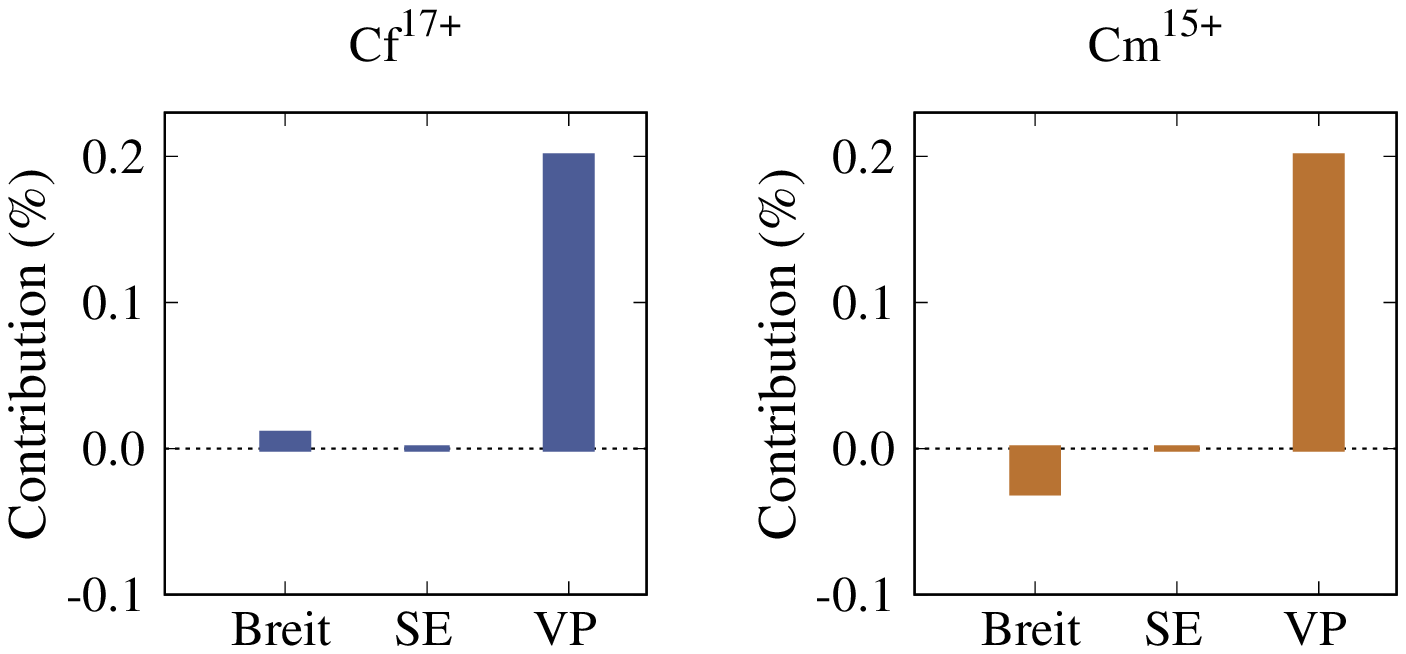}
\caption{Dominant contributions arising from Breit interaction, self-energy, 
	and vacuum-polarization to the sensitivity coefficient in the case 
	of $\rm Cf^{17+}$, $\rm Cm^{15+}$, and $\rm Bk^{16+}$.}
\label{sensitivity_bqed}
\end{figure*}

Fig. \ref{linear_q} shows the linear fit of $E$ vs. $x$ data for 
considered HCIs, which is used to extract the sensitivity 
coefficients for the proposed clock transitions. We further account for
the contributions arising from the Breit interaction, vacuum-
polarization effect, and self-energy correction to obtain the final recommended 
sensitivity coefficients, $q_{\rm final}$. These corrections are shown in the 
Fig. \ref{sensitivity_bqed}. The self-energy contribution is negligible 
for all the states considered, while the vacuum-polarization correction 
remains below 0.3\% across the three ions. Its largest contribution, 
approximately 0.3\%, occurs for the $6p_{3/2}$ state in $\rm Bk^{16+}$. 
In $\rm Cf^{17+}$, the vacuum-polarization correction is approximately 
0.2\% for the $6p_{1/2}$ state, while in $\rm Cm^{15+}$, it is approximately 
0.2\% for all the states considered. The Breit contribution is even smaller, 
with the largest corrections being $-0.009\%$, $-0.03\%$, and $-0.04\%$ for the 
$6p_{1/2}$ state in $\rm Cf^{17+}$, the $5f_{7/2}$ state in $\rm Cm^{15+}$, 
and the $5f_{7/2}$ state in $\rm Bk^{16+}$, respectively. 
Although these contributions are small relative to the dominant 
Coulomb-correlation effects, their consistent inclusion provides a more complete 
assessment of the sensitivity coefficients and strengthens the reliability of the 
recommended values for precision clock applications. 

\section{Theoretical uncertainty} 

Based on the approximations used in our calculations, we identify 
four primary sources which contribute to the uncertainties of calculated 
one-body matrix elements. The first source of uncertainty arises from the 
truncation of basis set. Our computed properties are observed to converge 
well with a negligible change on further augmentation in the basis. 
Hence, we neglect the uncertainty associated with this source.
The second contribution originates from truncating the dressed Hamiltonian 
to second order in $T^{(0)}$ in the evaluation of atomic properties \cite{mani-11,ravi-21b}. 
In our earlier investigations \cite{mani-10}, iterative inclusion of higher-order 
terms revealed that the contributions from third- and higher-order terms 
in $T^{(0)}$ are below 0.1\%. We, therefore, adopt 0.1\% as an upper bound 
associated with this source.
The third contribution arises from the partial treatment of triple excitations 
in FSRCC theory. Since perturbative triples account for dominant effects of 
triple excitations, the remaining higher-order contributions are expected to 
be relatively small. Among all transition properties computed in this work, the 
largest contribution of 0.3\% is observed in the case of lifetime of $6p_{1/2}$ 
state in $\rm Cf^{17+}$. Accordingly, we adopt 0.3\% as an upper bound arising 
from this source. 
The fourth contribution stems from the omission of frequency-dependent Breit 
interaction. In our previous study of Ra \cite{chattopadhyay-14}, a series of 
GRASP2K calculations indicated that this effect contributes less than 0.13\% to the 
calculated properties. Therefore, we take 0.13\% as an upper bound from 
this source. Additional sources of uncertainty, such as higher-order couplings 
involving vacuum-polarization and self-energy corrections, quadruply excited 
cluster amplitudes, and other neglected higher-order many-body effects, are 
expected to contribute at a significantly lower level. Their cumulative 
contribution is estimated to be below 0.1\%.
Combining all these upper bounds, the overall theoretical uncertainty in our 
calculated E1, E2, and M1 transition matrix elements is estimated to be 
less than 1\%.

Since the BBR shifts and magic rf-drive frequencies are derived from the 
calculated polarizabilities, uncertainties in these systematic shifts are 
governed primarily by the uncertainty in polarizability. 
To estimate this uncertainty, we refer to our previous work \cite{palki-26a}, 
in which we used a sum-over-states approach to calculate the scalar 
polarizabilities, $\alpha_s$, of the ground and excited clock states in 
$\rm Ca^{+}$ and $\rm Sr^{+}$. And, we estimated an 
uncertainty of approximately $2\%$ in $\alpha_s$. We therefore adopt $2\%$ 
as a conservative upper-bound estimate for the uncertainties in the 
calculated polarizabilities in the present work.

For the electric-quadrupole shift, the relevant quantity is the electric quadrupole 
moment of the clock state. Although a separate convergence analysis of the 
quadrupole moments is not presented, we have analyzed the convergence of
E2 transition matrix elements. Since both the quantities are determined by
same rank-2 electric-quadrupole operator \cite{jonsson_book}, the observed 
convergence of E2 matrix elements provides strong evidence for the reliability 
of our calculated quadrupole moments.


\bibliography{hci}